\documentclass[11pt,preprint]{aastex}
\usepackage{lscape}
\usepackage{graphicx}
\usepackage{indentfirst}
\usepackage{subfig}

\newcommand\ii{{\sc ii}}
\newcommand\iii{{\sc iii}}

\title{Chemical Abundances of Seven Irregular and Three Tidal Dwarf Galaxies in the M81 Group}
\author{Kevin V. Croxall, Liese van Zee}
\affil{Department of Astronomy, Indiana University, 727 East 3rd Street, Bloomington, IN 47405}
\email{kcroxall@astro.indiana.edu, vanzee@astro.indiana.edu}

\author{Henry Lee}
\affil{Gemini Observatory, Southern Operations Center, Casilla 603, La Serena, Chile}
\email{hlee@gemini.edu}

\author{Evan D. Skillman}
\affil{Astronomy Department, University of Minnesota, Minneapolis, MN 55455}
\email{skillman@astro.umn.edu}

\author{Janice C. Lee\altaffilmark{1}}
\affil{Observatories of the Carnegie Institution of Washington, 813 Santa Barbara Street, Pasadena, CA, 91101}
\altaffiltext{1}{Hubble Fellow}
\email{jlee@ociw.edu}

\author{St\'ephanie C\^ot\'e}
\affil{Canadian Gemini Office, Herzberg Institute of Astrophysics, National Reseach Council of Canada, 5071 West Saanich Road, Victoria, BC V9E 2E7, Canada}
\email{stephanie.cote@nrc-cnrc.gc.ca}

\author{Robert C. Kennicutt Jr}
\affil{Institute of Astronomy, University of Cambridge, Madingley Road, Cambridge CB3 0HA, UK}
\email{robk@ast.cam.ac.uk}
\and

\author{Bryan W. Miller}
\affil{Gemini Observatory, Southern Operations Center, Casilla 603, La Serena, Chile}
\email{bmiller@gemini.edu}

\begin{document}  
\begin{abstract}
 We have derived nebular abundances for 10 dwarf galaxies belonging to the M81 Group, including several galaxies which do not have abundances previously reported in the literature.  For each galaxy, multiple H~\ii\ regions were observed with GMOS-N at the Gemini Observatory in order to determine abundances of several elements (oxygen, nitrogen, sulfur, neon, and argon).  For seven galaxies, at least one H~\ii\ region had a detection of the temperature sensitive [OIII] $\lambda$4363 line, allowing a ``direct" determination of the oxygen abundance. No abundance gradients were detected in the targeted galaxies and the observed oxygen abundances are typically in agreement with the well known metallicity-luminosity relation.  However, three candidate ``tidal dwarf'' galaxies lie well off this relation, UGC~5336, Garland, and KDG~61.  The nature of these systems suggests that UGC 5336 and Garland are indeed recently formed systems, whereas KDG~61 is most likely a dwarf spheroidal galaxy which lies along the same line of sight as the M81 tidal debris field.  We propose that these H~\ii~regions formed from previously enriched gas which was stripped from nearby massive galaxies (e.g., NGC~3077 and M81) during a recent tidal interaction. 
\end{abstract} 
 
      
\section{Introduction}
Due to their low dust content, simple kinematics, modest to negligible abundance gradients, and prevalence in the universe, dwarf galaxies are excellent objects to use as probes of galaxy evolution (e.g., Hunter \& Gallagher 1985; Kobulnicky \& Skillman 1997).  
Additionally, these small galaxies may be remnants of the building blocks that combined to form larger galaxies early in the history of the universe (e.g., Bullock \& Johnston 2005).  Thus, understanding the chemical evolution of dwarf galaxies is critical for constraining galaxy evolution models.  Indeed, numerous studies have investigated dwarf galaxy abundances (e.g., Lequeux et al.\ 1979;  Skillman et al.\ 1989; Richer \& McCall 1995; Kunth \& \"Ostlin 2000 and references therein).  These studies have established the well known metallicity-luminosity relation, wherein more massive galaxies have higher oxygen abundances.  While this empirical trend has been measured repeatedly, its underlying cause(s) remains controversial.  On the one hand, greater retention of enriched gas by the deeper potential wells of more massive galaxies could produce the relation (e.g., Gibson \& Matteucci 1997).  However, another possibility is that pristine gas is processed more efficiently by larger galaxies.  Thus, a combination of gas transport (i.e., inflows and outflows) and systematic variations in star formation efficiency could produce this trend (e.g., Dalcanton 2007). 

At the same time, most galaxies do not exist in isolation.  Evolution and gas transport are influenced via galaxy-galaxy interactions; such influence is likely quite extreme in some cases.  If galaxies are hierarchically built structures, such interactions may be the {\sl de facto} method of galaxy evolution.  Thus, interacting galaxies may further inform us as to how the composition of the universe is evolving.  Has interaction significantly altered the chemical composition of systems?  The nearby M81 Group (D $\sim$ 4~Mpc) offers an ideal laboratory in which to investigate the signatures of chemical enrichment in a dynamic environment.  
This prominent group is one of the nearest galaxy associations to the Local Group.  Hence, even low mass group members may be observed and included in studies.  Additionally, the proximity of the M81 Group has permitted accurate distance determinations (Karachentsev et al.\ 2002 and references therein) and resolved studies of the stellar populations (e.g., Weisz et al.\ 2008).
As recently as 300 Myrs ago, the primary galaxies of the M81 Group experienced a dramatic collision (e.g., van der Hulst 1979; Yun et al.\ 1994).  The tidal HI debris still connects the three most massive galaxies involved in the event: M81, M82, and NGC 3077.  There have been detailed studies of this tidal debris in both HI and CO (Taylor et al. 2001, Chynoweth et al. 2008, Brinks et al. 2008).  Additionally, several clumps of new star formation, likely induced by the three-body interaction, have been identified (e.g., Davidge 2008).  

Indeed, it has been suggested that some dwarf galaxies within the group have recently been formed through tidal interactions (Makarova 2002;  Sabbi et al. 2008; Weisz et al.\ 2008).  Accordingly, one may hypothesize that the metallicity of such tidally formed galaxies could be elevated, compared to other galaxies of similar mass (e.g., Weilbacher et al. 2003); the gas from which they formed may be pre-enriched by the larger system from which the gas was stripped.  To investigate this possibilty, we have obtained spectra for 10 of the dwarf galaxies of the M81 Group, including some of the ``tidal dwarf" candidates, to determine the intrinsic metallicity of these systems. Several of these galaxies do not have previously determined abundances in the literature.  
 
 We present these data and discuss their implications for the evolution of the M81 Group.  In \S2 we present the observations and data processing.  The emission line measurements and abundance determinations are discussed in \S3 and \S4, respectively.  The results of the dwarf irregular galaxies are discussed in the context of the M81 Group in \S5, while tidal dwarf galaxies are discussed in \S6.  Lastly, \S7 summarizes our findings.  We adopt 12$+$log(O/H) = 8.93 (Anders \& Grevesse 1989) as the solar value of the oxygen abundance for the present discussion.

\section {Observations}
\subsection{Optical Spectroscopy}
Low-resolution spectra of ten M81 Group dwarfs were obtained\footnote[1]{Program ID GN-2006A-Q-26.}  in queue mode with the Gemini Multiple Object Spectrograph (GMOS) instrument (Hook et al. 2004) on Gemini North during May 2006.  Galaxies were selected to span a large range of both star formation rate and optical luminosity.  Global parameters for the selected targets are listed in Table~\ref{t:param}.  Masks were cut to place 5\arcsec\ long slitlets with a 1.5\arcsec\ slit-width on regions observed to emit significant flux in the H$\alpha$ pre-imaging observations. Blue spectra were obtained using the  B600 (600 lines mm$^{-1}$) diffraction grating with a resolution of 0.45 $\rm{\AA}$ pixel$^{-1}$.  Red spectra were obtained using the R600 diffraction grating with a resolution of 0.47 $\rm{\AA}$ pixel$^{-1}$.   Spectral observations were typically 2$\times$900s for both blue and red setups.  We used $4 \times 2$ spectral-spatial binning, respectively, resulting in a spectral resolution of 1.8 $\rm{\AA}$ per binned pixel.  Observations were centered at 4500 $\rm{\AA}$ and 4550 $\rm{\AA}$ in the blue, and 7000 $\rm{\AA}$ and 7050 $\rm{\AA}$ in the red, to ensure full spectral coverage across the two detector gaps.  We did not use an order-blocking filter for any of the spectra.  This yielded complete spectral coverage from roughly 3500 $\rm{\AA}$ to 8000 $\rm{\AA}$ for the majority of slitlets. 

The astrometric slit positions of emission sources presented in this work are listed in Table~\ref{t:positions}.  Five apertures revealed regions of strong hydrogen emission yet displayed no sign of oxygen lines;  these sources were not analyzed as part of this work as they do not seem to be H~\ii\ regions, but their positions are also listed in Table~\ref{t:positions}.  The position angle of the slitlets and the average parallactic angle for each pointing are also given in Table~\ref{t:positions}.  Due to scheduling constraints, it was not always possible to line up the slits with the parallactic angle to reduce the light loss from atmospheric refraction.   The fields most affected by this misalignment are UGC 4459, UGC 5139, UGC 5918, and UGC 8201.  

The spectra were reduced and analyzed using the IRAF\footnote[2]{IRAF is distributed by the National Optical Astronomical Obsevatories.} GMOS package.  Reduction of spectra included bias subtraction and flat fielding based on observations of the quartz halogen lamp on the GCAL unit.  A spectral trace of a bright continuum source was used to define the trace for all slits in each field. The shape of this trace was consistent in all exposures of a given field.  Wavelength calibration was obtained by observations of CuAr arc lamps interspersed between observations of the target galaxies.  

As all of the galaxies in this sample have relatively large angular extents, most filled a large fraction of the $5\arcmin \, \times \, 5\arcmin$ GMOS-N field of view.  Thus, the edges of each 5\arcsec\ slit lie well within the glow of the diffuse ionized medium of the host galaxy.  Furthermore, some H {\sc ii} regions are quite extended and completely fill these short 5\arcsec\ slitlets.  Therefore, we investigated whether standard local sky subtraction procedures could be used for these observations, given the concern that slits include both sky and diffuse ionized gas.  Indeed, local sky subtraction often over-subtracted strong lines relative to weaker emission lines, as expected since the diffuse ionized gas is at a different temperature than the compact H~\ii\ region.  Even in the most compact H {\sc ii} regions observed, a 10\% reduction in the flux of strong lines was found when the sky was measured locally.  

Thus, a more accurate sky background was measured in slits located away from the main body of the target galaxy.  Average red and blue sky spectra were created by combining multiple one dimensional spectra from slits sampling the sky for every galaxy observed.  
While the same physical region was extracted from both the blue and red setups, the size of this extraction region varied from slit to slit.  Thus, the amount of sky background in each slit was different, as it depended on the spatial extent of the slit with significant line emission.
 To ensure the proper level of sky background was subtracted, spectral regions populated by sky lines and free of any substantial extraterrestrial emission line were used as reference regions (e.g., 7150--7300 \AA).  Average sky spectra were scaled such that residuals in the reference regions were minimized when object and sky spectra were differenced.  Subsequently, the scaled backgrounds were subtracted from the one-dimensional H{\sc ii} region spectra.

One-dimensional spectra were corrected for atmospheric extinction and flux calibration based on observations of the flux standard Feige 67 (Oke 1990). The flux standard was observed on each night science data were obtained.  A representative flux-calibrated spectrum is shown for each target galaxy in Figure \ref{fig:spec}.  Note that red and blue continuum levels are in excellent agreement, despite observations being non-simultaneous.  This confirms stable sky conditions and indicates that the extraction regions were well matched in both the red and blue setups.  These example spectra are of similar quality to other spectra used for the abundance determinations of each target.  

\subsection{Line intensities}
The strength of emission features were measured in the one-dimensional spectra and subsequently corrected for Balmer absorption and line of sight reddening.  Reddening estimates were determined by comparing the ratios of flux from Balmer emission lines.  Ratios of intrinsic line strength were interpolated from the tables of Hummer \& Storey (1987) for case B recombination, assuming T$_e$=12500 K and N$_e$=100 cm$^{-3}$.  When the temperature sensitive [O {\sc iii}] $\lambda$4363 line was detected, temperatures derived from the [O {\sc iii}] lines were used rather than 12500 K.  The Galactic reddening law of Seaton (1979) parameterized by Howarth (1983) was applied to derive the reddening function normalized at H$\beta$, assuming R=A$_V$/E$_{B-V}$=3.1.  When the measured reddening coefficient, c$_{H\beta}$, differed for the H$\alpha$/H$\beta$ and H$\gamma$/H$\beta$ line ratios, an underlying Balmer absorption of up to 2${\rm \AA}$ was applied.  The determined values of c$_{H\beta}$ are reported in Table~\ref{t:lineratiob}.  

Reddening corrected line intensities measured from H {\sc ii}  regions in the target fields are reported in Tables~\ref{t:lineratiob} and \ref{t:lineratior}.  The regions are grouped according to the field with which they are associated, as listed in the first column.    The error associated with each measurement is determined from measurements of the Poisson noise in the line measurement, error associated with the sensitivity function, Poisson noise in the continuum, read noise, sky noise, flat fielding calibration error, error in continuum placement, and error in the determination of the reddening.   Note that while the spectral coverage permitted measurements of lines redward of the [Ar {\sc iii}] $\lambda$7136 line, the numerous skylines made measurements in the red portion of the spectra very uncertain.  
Furthermore, both [O {\sc i}] lines in our spectral region are coincident with strong atmospheric emission features.  Occasionally, the resulting profile of the sky-corrected [O {\sc i}] $\lambda$6300 and $\lambda$6364 lines had broad wings indicative of over subtraction.  Thus, measurements of [O {\sc i}] lines are not reported when sky subtraction did not yield trustworthy line profiles.  

\subsection{Diagnostic Diagrams}
Since the red and blue spectra were not obtained simultaneously and, further, queue scheduling did not always permit observations along the parallactic angle, diagnostic diagrams were examined to verify we recovered reliable line ratios, and hence reliable abundances.  For example, the [N~\ii]/H$\alpha$ and [O~\iii]/H$\beta$ sequence is shown in Figure \ref{fig:diag}a along with a theoretical curve from models of H~\ii\ regions (Baldwin, Phillips, \& Terlevich 1981).  H~\ii~regions from the current study are shown as circles and pentagons, while the triangles indicate H~\ii~regions from a study of spiral galaxies (van Zee et al.\ 1998).  Following McCall et al.\ (1985), Figure \ref{fig:diag}b shows the same diagnostic, replacing [O~\iii]/H$\beta$ with the sum of [O~\ii]/H$\beta$  and [O~\iii]/H$\beta$, commonly referred to as R$_{23}$.  Our data are in good agreement with both theoretical predictions and the general locus of galaxies in all diagnostic diagrams. 

Intriguingly, H~\ii~regions from the observed candidate tidal dwarfs, KDG 61, Garland, and UGC 5336, do not occupy the same parameter space as H~\ii~regions from the dwarf irregulars in our sample.  Rather, both diagnostic trends indicate they resemble H~\ii~regions observed in larger spiral galaxies.  We discuss these targets in more detail in \S5.

\section {Nebular Abundances}
Determination of elemental abundances from spectra of ionized gas requires determination of (i) the electron density (n$_e$), (ii) the electron temperature (T$_e$), and (iii) an estimate of the abundance of atomic species in unobserved ionic states.  
In spectra where the [O {\sc iii}] $\lambda$4363 line is detected, we may directly follow the methodology of Osterbrock \& Ferland (2006), since this line is highly sensitive to the electron temperature.  We adopt the standard practice of referring to this as the ``direct" method (Dinerstein 1990).  When measurement of the weak line is impossible, we have employed a strong line method wherein an oxygen abundance is deduced from photoionization models and measurements of the strong [O {\sc ii}] $\lambda$3727 and  [O {\sc iii}] $\lambda$5007,4969 lines (e.g., Edmunds \& Pagel 1984, McGaugh 1991).  Subsequently, a consistent electron temperature is deduced based on the strong-line oxygen abundance (van Zee et al.\ 1997).  While the absolute measurement of this ``semi-empirical" approach is less certain, it still provides robust relative abundances since relative abundances are less dependent on the electron temperature. 

\subsection{Direct Abundance Determinations}
Osterbrock \& Ferland (2006) carefully describe the procedure to determine the electron density and the electron temperature of an H~\ii\ region. We mention here only the most relevant points in the determination of T$_e$ and n$_e$.

The [S {\sc ii}] $\lambda$6717/6731 line ratio is sensitive to the electron density of the H {\sc ii} region.  The majority of H {\sc ii} regions studied here are within the low-density limit of this diagnostic ratio [$I(\lambda6717)/I\lambda6731)>1.35$].  Accordingly, an electron density of 100 cm$^{-3}$ was assumed for most H {\sc ii} regions.  In the six cases where the line ratio did not lie below the low-density limit, electron densities were calculated from the observed [S {\sc ii}] ratio using a version of the FIVEL program (De Robertis 1987).  

Detection of the [O~\iii] $\lambda$4363 line in a spectrum permits a direct determination of the electron temperature of the ionized gas. We used the [O~\iii] $\lambda$4363 line for a temperature measurement only when the S/N in the line exceeded 2.8.  Following this threshold for detection, we were able to measure reliable reddening corrected [O~\iii] $\lambda$4363 fluxes for 26 H~\ii~regions (56\% of the sample).
These 26 H~\ii\ regions were distributed among 7 of the 10 observed fields.  
From these fluxes, electron temperatures were determined for the O$^{++}$ region of the nebula.  
Subsequently, the electron temperature in the O$^+$ region was calculated from the O$^{++}$ electron temperature using the approximation of Pagel et al.\ (1992).  With electron densities and temperatures in hand, emissivity coefficients for detected emission lines were determined using the FIVEL program (De Robertis et al. 1987).

While emission lines were detected for all dominant ionization states of oxygen, derivation of abundances for other elements requires accounting for the presence of atomic species whose ionization states were unobserved.  Nitrogen abundances were derived under the assumption N/O=N$^+$/O$^+$;  neon abundances were derived under the assumption Ne/O = Ne$^{++}$/O$^{++}$ (Peimbert \& Costero 1969).  In the cases of both sulfur and argon, the analytical ionization correction factors of Thuan, Izotov, \& Lipovetsky (1995) have been adopted.  When the [S~\iii] $\lambda$6312 line was not detected, the sulfur abundance was deemed to be too uncertain to report. 

\subsection{Semi-empirical Abundance Determinations}
In the 20 H~\ii\ regions where the [O~\iii] $\lambda$4363 line could not be cleanly measured or was simply too weak to be detected, we are unable to determine the electron temperature via a direct approach.  However, the strong oxygen lines ([O~\ii] $\lambda$3729 and [O~\iii] $\lambda$4959,$\lambda$5007) are sensitive to both oxygen abundance and electron temperature.  Therefore, a semi-empirical approach was used, where oxygen abundances were estimated from the strong line ratio (e.g., McGaugh 1991); we uniformly adopt an error of 0.20 for abundances determined via the semi-empirical method as discussed in van Zee et al.\ (2006).  In this approach, where photoionization models are used as calibration, the geometry of the H~\ii\ region is represented by the average ionization parameter, \=U, the ratio of ionizing photon density to particle density, as determined by the ratio of [O~\iii] to [O~\ii].  This additional parameter increases the spread of abundance estimates for a given R$_{23}$.  

The theoretical photoionization models of McGaugh (1991) are shown in Figure \ref{fig:mcg} along with the observed line ratios presented in this work; solid points indicate the [O~\iii] $\lambda$4363 line was detected while open points denote spectra where the line was unmeasurable.  In general, our data lie in the region predicted by theoretical models; however, three points fall off the model grid.  Two of these regions, KDG 61-9 and Garland-5, have detections of the [O {\sc iii}] $\lambda$4363 line; therefore, the electron temperature was determined directly from line ratios and does not rely upon photoionization models.  
The remaining point, UGC 5336-11, is from a supernova remnant.  As such, it would not be expected to follow predictions of pure photoionization but must be compared to models of shock ionization (see the discussion on UGC 5336 in \S4).

The strong-line ratio of ([O~\ii]+[O~\iii])/H$\beta$, or $R_{23}$, is a smooth function that depends on the electron temperature and oxygen abundance of a gas; however, $R_{23}$ is double valued.  Collisionally excited Lyman series emission accounts for the cooling of gas in a very low metallicity H~\ii\ region.  As the metal content increases, contributions to cooling from infrared fine structure lines become more important.  Around an oxygen abundance of one-third of the solar value (12$+$log(O/H) $\sim$ 8.4), cooling becomes dominated by fine structure lines, causing the R$_{23}$ surface to fold over despite the increased metallicity.  The fold in the $R_{23}$ surface is shown in Figure \ref{fig:mcg}.  The fold-degeneracy may be broken using the [N~\ii] to [O~\ii] line ratio as a diagnostic (e.g., Alloin et al.\ 1979).  In the majority of cases here, H~\ii\ regions exhibited log([N~\ii]/[O~\ii])$<-1.0$.  Accordingly, the ``lower-branch'' and thus lower values for the oxygen abundance were adopted for those H~\ii\ regions.  The high abundance value was generally adopted for regions in the ``upper-branch'' with log([N~\ii]/[O~\ii])$>-1.0$.  However, some ($\sim5$) of the observed H~\ii\ regions fell near this critical value.  In such instances, the two possible strong-line abundances were compared to abundances from other H~\ii\ regions within the field.  As the interstellar-medium chemical abundances of dwarf galaxies are generally spatially homogeneous (e.g., Kobulnicky \& Skillman 1997; Lee \& Skillman 2004; Lee et al.\ 2005, 2006), the abundance branch yielding a more consistent value with other spectra was adopted; often these other spectra had detections of $\lambda$4363, eliminating this uncertainty on their behalf. 

The selected H~\ii\ regions in UGC 5692 were all faint targets with large errors in the [N~\ii] to [O~\ii] ratio.  Thus, this diagnostic line ratio did not present a clear resolution to the R$_{23}$ degeneracy.  
While no spectra from UGC 5692 yielded a solid [O~\iii] $\lambda$4363 detection, two of the apertures had weak detections of this line. These detections were more consistent with higher temperature H~\ii\ regions.  Furthermore, the temperatures derived from the [O~\iii] line ratios yielded oxygen abundances consistent with the lower branch of the R$_{23}$ relation.  Therefore, the lower abundance was adopted for H~\ii\ regions in UGC 5692.

The strong-line abundance calibrations we have used are derived from zero-age H~\ii\ regions (McGaugh 1991).  As an H~\ii\ region ages, the characteristic spectrum and ionization parameter may evolve, introducing systematic abundance errors (e.g., Stasi\'nska \& Leitherer 1996; Olofsson 1997).
van Zee et al.\ (2006) showed that such corrections may be significant when log([O~\iii]/[O~\ii])$<-0.4$ in an H~\ii\ region.  Within the present sample, two regions in Garland (Garland-6 and Garland-9) and three regions in UGC 5336 (UGC5336-3, UGC5336-11, and UGC5336-12) lie within this highly uncertain parameter space.  Weisz et al. (2008) have shown concentrations of blue stars in both of these systems, indicative of star formation in the last 100 Myr.  Thus, these HII regions are associated with recent star formation, but we have no constraint on the character of the radiation field (i.e., reflective of a fully populated zero-age main sequence versus an evolved population of lower mass main sequence stars).  Lacking more information on the H~\ii\ regions sampled by Garland-6, Garland-9, and UGC5336-3, we have made no corrections for the age of the nebulae in the current analysis.  Abundance determinations from UGC5336-11 and UGC5336-12 are discussed further in \S4.  

\subsection{Abundances}
The abundance determinations for each H {\sc ii} region are presented in Table~\ref{t:abund}.  Abundances were calculated using the measured line strengths and corresponding emissivity coefficients as determined by FIVEL (De Robertis et al.\ 1987).  Errors associated with the derived abundances were clearly dominated by the uncertainty of the electron temperature.  The abundances of nitrogen and the $\alpha$-elements (Ne, S, and Ar) relative to oxygen are shown in Figure \ref{fig:alpha}, along with solar values from Anders \& Gervasse (1989).  At higher oxygen abundance, H~\ii\ regions do show an increased N/O ratio.  This relative increase is consistent with secondary production of nitrogen dominating the primary component at higher metallicities (see Figure 4 of Vila-Costas \& Edmunds 1993).  As expected for primary elements, $\alpha$/O trends appear to be constant as oxygen abundance increases.  The mean log (Ne/O) is $-0.82\pm0.12$; the mean log (Ar/O) is $-2.21\pm0.13$; the mean log (S/O) is $-1.50\pm0.10$.  These mean values are in agreement with large samples of H {\sc ii} regions in dwarf galaxies (e.g., Thuan et al.\ 1995; Izotov et al.\ 2004; Lee et al.\ 2004; van Zee et al.\ 2006). 

By placing multiple apertures on targeted galaxies, we are able to investigate the uniformity of oxygen abundance across dwarf galaxies in the M81 Group.   Except for the candidate tidal dwarfs, the standard deviation of oxygen abundances within each galaxy is smaller than the error associated with the individual measurements.  This reaffirms previous results which have shown no detectable abundance gradients within dwarf galaxies (e.g., Kobulnicky \& Skillman 1996, 1997).  While the suspected tidal dwarf galaxies Garland and UGC 5336 may show real spatial trends in oxygen enrichment, these galaxies are not representative of undisturbed dwarf galaxies.  These dwarf galaxies and their origins are discussed further in \S6.1~--~\S6.3.

The average oxygen abundance and log(N/O) for a given galaxy are tabulated in Table~\ref{t:param}.  The average abundance was determined with a weighted average of the individual measurements. This relies upon the assumption that each H~\ii\ region is an independent  measure of the composition of the galaxy, which is taken to be uniform.  Abundances determined using the direct method are more certain.  Thus, when direct and semiempirical determinations were available, the strong line abundances were not used in the average.  It should be noted that all strong line abundances agree, within the errors, with the average abundances determined for their respective galaxies.  Also, due to their larger uncertainties, the semi-empirical abundance determinations would have very little effect in weighted averages.

\section {Comparison with Previous Work}
Previous studies of the M81 Group have reported abundances for half of the galaxies in this sample.  We compare our findings with those of previous studies below.  In general, we note that the new results presented here are consistent with previous observations of these systems.  

{\it UGC 4459.} -- Hunter \& Gallagher (1985), Skillman, Kennicutt \& Hodge (1989), Hunter \& Hoffman (1999), Pustilnik et al.\ (2003) and Saviane et al.\ (2008) all obtained spectra of UGC 4459.  They determined an oxygen abundance, 12+log(O/H), of 8.59, 7.62, 7.79, 7.52$\pm$0.08 and 7.83, respectively, where [O~\iii] $\lambda$4363 was only detected by Pustilnik et al.\ (2003).  While the values of Skillman, Kennicutt \& Hodge (1989) and Pustilnik et al.\ (2003) are lower than the value determined here using the direct method (7.82$\pm0.09$), they are in excellent agreement with the semi-empirical abundance we calculate for UGC 4459 (7.61$\pm0.20$).  We adopt the mean oxygen  abundance determined from the direct method in this paper, which is in excellent agreement with both the Hunter \& Hoffman (1999) and Saviane et al.\ (2008) results.  

{\it UGC 5139.} -- A strong line abundance is reported for UGC 5139 based only on the detection of $\lambda$4959 and $\lambda$5007 (Miller \& Hodge 1996).  To derive an abundance, they estimated the strength of [O~\ii] based upon theoretical ratios of [O~\iii]/[O~\ii].  Our semi-empirical result for this same H~\ii\ region (7.75$\pm0.20$) is in excellent agreement with their value of 12+log(O/H)$= 7.7\pm0.3$.  Here, we adopt our direct-method average abundance of 8.00$\pm0.10$.

{\it UGC 4305.} -- Oxygen abundances of 12+log(O/H)=8.55, 7.92, and 7.71$\pm$0.13 were reported by Hunter \& Gallagher (1985), Masegosa et al.\ (1991), and Lee et al.\ (2003), respectively.  While the strong line abundance of Hunter \& Gallagher (1985) appears to have been placed on the ``upper-branch" of the strong line abundance models, the three HII regions where the [OIII] $\lambda$4363 line was detected by Masegosa et al.\ (1991) are in excellent agreement with our derived value of 7.92$\pm$0.10.  We note the presence of strong He \ii\ $\lambda$4686 emission in UGC 4305-11, which may indicate the presence of another ionizing source, such as a Wolf-Rayet star.  Such a composite spectrum could feature blends of lines which would not follow the general properties of a photoionized region.  Thus, we do not include UGC 4305-11 in our average abundance of UGC 4305.

{\it UGC 5666.} -- Multiple H~\ii\ regions in UGC 5666 were studied by Masegosa et al.\ (1991) and Miller \& Hodge (1996), yielding 12 + log(O/H) = 8.09 and 8.06, respectively; these results are from a combination of direct and empirical methods.  While the current results are very tightly clumped together, with a resultant average of 7.93 $\pm$ 0.05, they lie $\sim0.1$~dex below the previously reported abundances.   UGC 5666 extends well beyond the $5\arcmin \, \times \, 5\arcmin$ field of view of GMOS-N.  Therefore, diffuse galactic emission in the sky slit could have produced a systematic offset as discussed in \S2.1.  However, it is also worth noting Masegosa et al.\ (1991) obtained their data in 1984, before linear CCDs were widely employed, and thus a discrepancy of only 0.1 dex may not be significant.  

{\it UGC 5336.} -- An investigation of the optical counterpart of an X-ray source led to the acquisition of spectra from a U-shaped object north-east of UGC 5336 (Miller 1995).  Two of our apertures located in the UGC 5336 field lie upon this object.  While Miller (1995) reported anomalous H$\alpha$/H$\beta$ ratios for these regions and thus did not correct for line-of-sight reddening, our new observations yield normal H$\alpha$/H$\beta$ ratios indicative of foreground reddening.  In agreement with Miller (1995), we find the [S~\ii] $\lambda$6717,6731/H$\alpha$ for slits UGC 5336-10 and UGC 5336-11 are 0.6 and 0.7 respectively, clearly above the criterion necessary for an object to be classified as a supernova remnant (Skillman 1985; Smith et al.\ 1993).  

Since UGC 5336-10 and UGC 5336-11 are not from purely photoionized H~\ii\ regions, proper abundance determinations require shock models.  Thus, diagnostic plots using [N~\ii ]/H$\alpha$, [S~\ii] $\lambda$6731/H$\alpha$ and [O~\iii]/H$\beta$ have been constructed using the MAPPINGS III radiative shock ionization models (Allen et al.\ 2008).  Under the assumption of a simple low density shock model, both diagnostics indicate 12+log(O/H)$\sim8.4$.  Given that this value is consistent with the oxygen abundance results of photoionization models, we adopt the semi-empirical abundance determinations for these apertures.  The abundance determined from apertures sampling the supernova remnant (8.44$\pm0.20$) as well as UGC 5336 (8.86$\pm0.20$) are in excess of the findings of Miller (1995: 12+log(O/H)$\sim$8.0).  While these two objects may not be physically associated, they appear to be linked by a common H I envelope.  Accordingly, we include measurements from all four apertures in the abundance determinations for UGC 5336.

\section{The Metallicity-Luminosity Relation of Dwarf Irregulars in the M81 Group}
As the M81 Group is relatively nearby ($D \sim 4$ Mpc), we are able to perform detailed studies of the low mass dwarfs in addition to the more massive galaxies.  Thus, we may investigate the metallicity-luminosity relation of galaxies in the group over a large dynamic range in galaxy luminosity.  With the addition of abundances reported in this paper, oxygen abundances are now available for almost all of the gas rich M81 Group galaxies.  Table~\ref{t:m81} lists the gas-rich galaxies of the M81 Group from the compilation of Karachentsev (2005); for completeness, we include UGC~5918 in this table despite the fact that revised distance estimates place it behind the main M81 Group (Karachentsev 2002; 2005).  Tabulated errors in M$_B$ were determined based on estimates of uncertainties in photometry and distance determinations.  For galaxies whose distances were determined via the tip of the red giant branch method, we adopt a 10\%\ uncertainty in the distance (Karachentsev et al.\ 2002, 2004).  Errors of 0.34 and 0.4 Mpc were used for M81 and NGC 2403 respectively (Freedman et al.\ 1994; Freedman \& Madore 1988).  The M81 distance error was also adopted for UGC 5336, as it is in the tidal field of M81.  As the distance to UGC 5918 is based upon photometry of three bright stars (Sharina et al.\ 1999), we adopt an error of 30\%\ in its distance determination.  An error of 0.20~mag was adopted for galaxy magnitudes (Makarova 1999). 

 We show the relation between oxygen abundance and absolute blue magnitude for group members in Figure \ref{fig:metlum} along with the metallicity-luminosity relation of local galaxies, 12~+~log(O/H)~= $-$0.151M$_B$~+~5.67 (van Zee et al.\ 2006). Galaxies at both the high and low ends of the luminosity range of the M81 Group tend to follow the trend established in local field galaxies. However, the intermediate luminosity galaxies (UGC 4305, UGC 5666, and NGC 2366) seem to fall below this empirical relation.  In general, offsets from the global trend can be the result of different distance scales, systematic effects, or genuine galaxy evolution.  Here, the first is unlikely, as most distances for this sample are measured via reliable techniques, such as fitting the tip of the red giant branch or cepheid variability.   Further, while two of these three galaxies are the largest galaxies for which we obtained spectra on GMOS-N, and thus are most likely to be biased by the sky subtraction issues outlined in \S2.1, independent determinations of oxygen abundances in these galaxies also fall below the trend (Masegosa et al.\ 1991; Miller \& Hodge 1996).  Thus, this flattened slope may be the result of genuine evolution within the galaxies observed.  Indeed, this evolution could come from either luminosity enhancements or gaseous infall.

Specifically, the under-abundance (or over-luminosity) of oxygen in these intermediate mass galaxies could be attributed to special circumstances within these particular galaxies.  For example, all three of the galaxies in question are undergoing starbursts and show elevated star formation rates relative to their averages in the last 0.1 to 1 Gyr (Weisz et al. 2008).  The presence of bright star forming regions enhances the blue luminosity of a galaxy, relative to galaxies not currently undergoing large amounts of star formation.  In the case of a global starburst, the luminosity enhancement in B-band could be as much as 1-2 magnitudes.  For example,  a starbust with a star formation rate that is 10 times the average past rate for several hundred million years  will have a maximum luminosity enhancement of 1.5 magnitudes (van Zee 2001).  To investigate the impact of current star formation on the observed blue luminosity, we masked bright H~\ii\ regions in B-band images of UGC 5666 and NGC 2366, which provides a minimal estimate of the luminosity enhancement from current star formation.  Surface brightness profiles were measured in both the masked and original data.  For both of these galaxies, the luminosity enhancement due to star forming regions appears to be approximately 0.3 magnitudes.  Thus, the observed offset of even the most extreme outlier, UGC 5666, could possibly be explained if a global burst has dramatically increased its B-band luminosity, particularly given the scatter in the relation when plotted versus mass rather than luminosity (e.g., Tremonti et al.\ 2004).

Alternatively, rather than being over luminous, these specific galaxies could be under-abundant.  Intermediate mass galaxies are likely to possess deeper potential wells than the low mass counterparts that account for the majority of galaxies in a group.  Therefore, while these galaxies are somewhat removed from the interacting triple (M81, M82, and NGC 3077), it is plausible that gas has been stripped from other M81 Group members and fallen into these galaxies, thereby diluting the abundance of the interstellar gas in these galaxies.  However, archival 21 cm VLA observations reveal no obvious signs of tidal disturbance in these galaxies. Therefore, significant dilution by gaseous infall cannot have occurred  recently.

Conversely,  the apparent shallow slope may simply be an artifact of fitting a small number of galaxies.  Scatter is clearly evident in metallicity-luminosity relations (e.g., Tremonti et al.\ 2004, Lee et al. 2006) and this may simply be a case where the relation is not well sampled.  Indeed, if we broaden group membership to include all galaxies noted as members by Karachentsev et al.\ (2002), we find the slope and intercept of M81 Group galaxies agree within the errors to trends found for local galaxies (van Zee et al.\ 2006) and field dwarf irregulars (Lee et al.\ 2003), excluding the outliers UGC 5336, KDG 61, and Garland.
While it may be argued that a field relation is recovered here simply because we are including field galaxies which happen to lie near the M81 Group, this expanded sample includes galaxies which lie in an overdense region of the Local Supercluster, and are not just a random sample.  Furthermore, it is difficult to build a large sample for any group without extending the membership criteria; for instance, the majority of gas rich galaxies in the M81 Group that do not already have a measured oxygen abundance are extremely low luminosity systems with very low star formation rates (Table \ref{t:m81}).  Therefore, spectroscopic observations of these targets with current instruments are unlikely to yield reliable abundances due to their intrinsically faint emission lines.  Moreover, ionized gas in the vicinity of these low mass galaxies may not belong to dwarf galaxies in question.  All of these low mass galaxies lie in close proximity to M81, where neutral HI shows it has been clearly tidally disturbed.  Ergo, these galaxies may not  actually be  ``gas rich," and emission line abundances of HII regions in these fields may be representative of the tidal stream and not representative of the low mass galaxy (see \S6.1 for further evidence of this hypothesis).  Hence, to acquire a statistically significant sample the membership criteria must be relaxed to some degree.

Still, a physical difference in the metallicity-luminosity relation in a group environment when compared to isolated galaxies cannot be ruled out. Gas transfer among group members may have divided the M81 Group into two abundance regimes, 12~+~log(O/H) $\sim$ 7.9 and 8.7 (see Figure~\ref{fig:metlum}).  High abundance H~\ii\ regions may correspond to those regions where gas is enriched by stellar evolution in massive galaxies; low abundance H~\ii~regions may all share a common composition due to gas infall that has diluted more massive irregular galaxies.  The abundance difference between massive galaxies and low mass galaxies of the M81 Group is not extreme; could such an effect in group abundances account for some of the observed scatter in the metallicity-luminosity relation?  

\section{The M81 Group Tidal Dwarfs}
One possible consequence of a galaxy interaction is the creation of new systems, so called tidal dwarfs.  
Given this formation mechanism, tidal dwarfs are expected to have little to no dark matter and to have chemical enrichment histories related to those of the parent galaxy (e.g., Duc et al.\ 2000).  Similar to other candidate tidal dwarf galaxies (Duc \& Mirabel 1998),  H {\sc ii} regions in and near KDG 61, UGC 5336, and Garland exhibit enhanced metallicities compared to other galaxies at similar luminosities.  Furthermore, these three dwarfs all lie within the tidal bridges of neutral hydrogen connecting M81, M82, and NGC 3077, as illustrated in Figure \ref{fig:map}.  This alignment has led to the suggestion that these dwarfs formed recently, as a result of tidal interactions within the group.  Notably, the oxygen abundances measured in these fields are similar to abundances measured in M81 and NGC 3077, as indicated by the dashed lines in Figure \ref{fig:metlum}.  We discuss these three galaxies in more detail below.

\subsection{KDG 61}
KDG 61 was classified as a dwarf irregular based on the presence of large amounts of neutral hydrogen and a central region of ionized gas, despite having the global $B\!-\!V$ color of a dwarf spheroidal (Johnson et al. 1997).  However, KDG 61 clearly lies in projection with the tidal stream connecting M81 and NGC 3077 (see Figure \ref{fig:map}; Yun 1994).  Thus, it is uncertain if neutral and ionized gases are in fact associated with the stellar component.  Only one slitlet in the present study contained emission from an H~\ii\ region.  Furthermore, as shown in Figure \ref{fig:kdg}, no other regions of H$\alpha$ emission are detectable in the field, even with a deep H$\alpha$ exposure (Croxall \& van Zee 2009, in preparation).  Therefore, comparison of this enhanced abundance with measurements from other H {\sc ii} regions in this field is not feasible.  

It is possible that the object we observe as KDG 61-9 is, in fact, a chance alignment between a dwarf spheroidal galaxy and an unassociated H~\ii\ region.  The distance to KDG 61 has been determined via fitting the tip of the red giant branch, which is independent of velocity measurements of both the gaseous and stellar content in the field.  Therefore, we can use the kinematics of the system to investigate the associations of observed components.  To explore the kinematics of KDG 61, we use both the multi-slit data and a higher spectral resolution long slit spectrum available in the Gemini Science Archive\footnote[3]{Program ID GN-2006B-Q-29.}.  As illustrated in Figure \ref{fig:kdg}, the radial velocities of the observed optical emission lines are consistent with the H{\sc i} velocity field (Yun et al.\ 1994).  Unfortunately, the strong magnesium absorption lines at 5175 and 5269 \AA\ that are often used to determine stellar velocities in dwarf spheroidal galaxies were not detected in either the multi-slit data or in the diffuse body of KDG 61 in the long slit data.  Therefore, we cannot say if the stellar component of KDG 61 is also moving with the tidal stream. However, we note that H$\beta$ in absorption was detected in the central high surface brightness region, A 0951+68 A, at a different velocity.  Thus, either this second high surface brightness object is a foreground star, and not a part of KDG~61, or the stellar component of the dwarf galaxy KDG 61 has a different velocity than the emission regions.  If the latter, then KDG 61-9 is not associated with KDG 61.

The emission seen in KDG 61-9 is produced by a remarkably hard ionization field (see Figure~\ref{fig:spec}).  Hence, it may not be a H~\ii\ region as we have presumed.  Indeed, the physical shape of an expanding bubble (Figure \ref{fig:kdg}), paired with the hard radiation field may indicate this object is a supernova remnant, a planetary nebula, or a wind-blown bubble.  Supernova remnants exhibit strong [N~\ii ] and [S~\ii] emission relative to H$\alpha$. Yet KDG 61-9 exhibits an [S~\ii] $\lambda$6717,6731/H$\alpha$ of 0.096$\pm$0.070, well below the accepted value of 0.4 for classification as a supernova remnant (Smith et al.\ 1993).  Planetary nebula are easily produced by an old stellar population and could contain substantial self enrichment (e.g., Pe\~{n}a et al.\ 2007).  However, at the distance of the M81 Group, the angular extent of the observed emission indicates a physical size on the order of 75~pc ($\sim4\arcsec$).  This is clearly too large for a typical planetary nebula, which would be unresolved if associated with KDG 61.  Furthermore, we detect He \ii~$\lambda$4686 line, which may indicate a Wolf-Rayet-Star which has created a wind-blown bubble.  We note the the He~\ii\ line is very concentrated in the two dimensional spectra of KDG 61-9; also, it is not apparent in either the spectrum of Johnson et al.\ (1997) or the long slit Gemini data.  The size of this feature is consistent with the approximate size of bubbles found in the LMC (50 $-$ 100 pc, Chu \& Mac Low 1990).  
While the nature of this emission is unclear, the position and velocity agreement with the neutral hydrogen of the M81 tidal stream seem rather coincidental for a foreground object.  Therefore, while we cannot eliminate other possibilities, an H~\ii\ region associated with the interaction of NGC 3077 and M81 seems the most likely  source of this emission.  Regardless, the detected emission seen towards  KDG 61 seems enriched beyond what is expected for a dwarf galaxy of such luminosity.

\subsection{UGC 5336 (Ho {\sc ix})}
The origins of UGC 5336, which is located just to the east of its massive neighbor M81,  have been the subject of debate for some time (e.g., Karachentseva et al.\ 1985, Makarova 2002). Recent work has highlighted the possible recent origins of UGC 5336.  While a dense concentration of blue stars is found in UGC 5336, the few red stars found to exist in the field are likely associated with the outskirts of M81 (Sabbi et al.\ 2008).  The detailed star formation history further indicates a large fraction of the stellar mass has been formed within the past 200 Myr (Weisz et al.\ 2008).  In this study, spectra of two H~\ii~regions and two regions of a supernova remnant in UGC 5336 were obtained.

All apertures which contained emission associated with the large region of HI in which UGC~5336 is found indicate a large enhancement of oxygen.  Unfortunately, these spectra do not show any indication of the temperature sensitive [O~\iii] $\lambda$4363 line.  This non-detection may be the result of low S/N or due to the fact that [O~\iii] $\lambda$4363 becomes more challenging to detect as the metallicity increases.   Therefore, the oxygen abundance is uncertain due to the double valued nature of the strong line abundance indicators.  Indeed, the lower branch of R$_{23}$ models yield 12+log(O/H)$\sim8.0$, close to the prediction for a galaxy with the luminosity of UGC 5336.  However, the observed [N~\ii]/[O~\ii] ratios place it clearly in the high metallicity regime.  The [NII]/H$\alpha$ ratio also provides constraints since it has been shown to correlate with oxygen abundance (e.g., van Zee et al.\ 1998; Denicol\'o et al.\ 2002; Salzer et al.\ 2005).  While this indicator is not particularly accurate ($\pm0.3$ dex), it does provide an indication upon which branch of the R$_{23}$ relation an H~\ii\ region falls.  Using the relation 12 + log(O/H) = 1.02log([N~\ii]/H$\alpha$) + 9.36, from van Zee et al.\ (1998), yields 12+log(O/H)=8.7$\pm$0.2 for both H\ii\ regions in UGC 5336.  Additionally, shock ionization models for  gas with 12+log(O/H)=8.0 are well outside the errors of line ratios from the SNRs UGC 5336-10 and UGC 5336-11.  Thus, while higher signal-to-noise ratio spectra would be desirable, the gas seems to be genuinely enriched.

Our derived oxygen abundance for UGC 5336 supports a scenario wherein the recent star formation occurred using gas stripped from the outskirts of M81.  As shown in Figure \ref{fig:metlum}, the derived oxygen abundance of UGC 5336 is similar to outer H~\ii~regions in M81.  Furthermore, the envelope of neutral hydrogen around UGC 5336 is directly linked to the gas of M81 (see Figure \ref{fig:map}).
Hence, the high metallicity is indicative of the presence of pre-enriched gas in the ionized medium of UGC 5336.  While this is not conclusive proof that UGC 5336 is entirely the result of the recent dynamic upheaval, it clearly shows the chemical evolution of the galaxy was severely affected by the interactions of its massive neighbors.

\subsection{Garland}	
Garland was noted by Karachentsev et al.\ (1985) as a separate complex near NGC 3077.  Similar to KDG 61 and UGC 5336, Garland clearly lies in the tidal stream created during the recent interaction (see Figure \ref {fig:map}).  Oxygen abundances determined from ionized gas scattered throughout this region yield a wide range of metallicity (Figure \ref{fig:metlum}).  Line diagnostics all indicate that these H{\sc ii} regions belong on the upper branch of photoionization models, though lower branch photoionization models cannot be completely ruled out.  Furthermore, Garland-5 has a solid detection of the [O {\sc iii}] $\lambda$4363 line (S/N~$\sim$~4.5), confirming the presence of enriched gas in this debris littered region.

As with UGC 5336, spatial maps of the stellar density in Garland depict clumps of recently formed blue stars with a paucity of an old red population (Weisz et al.\ 2008).  Additionally, the old red population is distributed smoothly as would be expected for stars associated with the larger galaxies.  Given the star formation history of Garland, the episode of recently-induced star formation coincides with the nearest approach between M81 and NGC 3077, $\sim250$~Myr ago (Yun et al.\ 1999; Weisz et al.\ 2008).  Indeed, the H$\alpha$ map of the field seems to show tidal tails, which are known to result from galaxy encounters (Croxall \& van Zee 2009, in preparation).  While a low mass dwarf galaxy may exist among the numerous clumps of H$\alpha$ emission, tidally stripped gas likely accounts for the majority of star forming knots.

\section {Conclusions}
Spectroscopic observations acquired with GMOS-N were used to determine chemical abundances for 46 H~\ii\ regions in ten dwarf galaxies belonging to the M81 Group.  Oxygen, nitrogen, sulfur, neon, and argon abundances were derived and presented in the context of the evolution of this group of galaxies.  The oxygen abundances for intermediate mass galaxies in the M81 Group seem to lie below the metallicity-luminosity relation observed among other dwarf galaxies.  While systematic effects may play a role in this offset, data from multiple studies indicate these galaxies have lower than expected oxygen.   NGC 2403, UGC 4305, and UGC 5666 are all undergoing a burst in star formation. Perhaps galaxies undergoing a starburst may not be expected to follow the same trend as more quiescent galaxies.  Alternatively, this may be an indication that evolution in a group environment differs from the evolution of isolated galaxies, as traced by the metallicity-luminosity relation. However, the small number of galaxies in the M81 Group may cause these data to appear more significant than they would seem in a large sample.  

Additionally, we have identified three locations within the M81-M82-NGC 3077 tidal stream where enriched gas is found external to a massive galaxy.  Two of these objects, UGC 5336 and Garland, are candidate tidal dwarf galaxies. The third galaxy, KDG 61, appears to have been misclassified as a dwarf irregular;  in this instance, we propose the H~\ii\ emission and neutral hydrogen belong to tidal bridges within the group and not to KDG 61.  In all cases, the chemical enrichment is consistent with abundances found in the outskirts of the massive galaxies M81 and NGC 3077, located near the target dwarfs (Zaritsky et al.\ 1994; Storchi-Bergmann et al.\ 1994).  As these galaxies were involved in the recent interaction, this may indicate that these regions are simply tidally stripped gas and not associated with the stellar components of the dwarf galaxies.  Indeed, two of these locations seem tidally unstable, following the criteria of Karachentsev et al.\ (2004), while the third seems likely to be a chance alignment of a dwarf spheroidal galaxy and an H{\sc ii} region in the tidal stream.  

If the enhanced abundances of these fields are indeed due to the presence of pre-enriched gas, then we expect to find other enriched clumps in the M81 tidal stream.  Indeed, enriched pockets of gas should exist throughout the entire debris system.  The measurement of oxygen abundances spanning this tidal system would place excellent constraints on models of gas mixing subsequent to interactions.  The numerous H~\ii\ regions in the fields of UGC 5336 and Garland hint at a spread in oxygen abundance. If verified, such abundance patterns may give insight into the formation of tidal systems.  Could gas in tidal tails be traced to the different host galaxies via metallicity determinations?  Are there analogous systems which have undergone interactions and show enriched abundances in their own tidal streams?  It is vital to understand how encounters and collisions affect the chemical and evolutionary histories of systems, if we are to understand groups at cosmological redshifts where the rate of galaxy encounters rises substantially. 

\acknowledgments
K.V.C. would like to thank thank John Salzer for insightful discussions about the spectra presented here.  This research was supported by NSF grant AST05-06054 and based on observations obtained at the Gemini Observatory, which is operated by the Association of Universities for Research in Astronomy, Inc., under a cooperative agreement with the NSF on behalf of the Gemini partnership: the National Science Foundation (United States), the Science and Technology Facilities Council (United Kingdom), the National Research Council (Canada), CONICYT (Chile), the Australian Research Council (Australia), Minist\'{e}rio da Ci\^{e}ncia e Tecnologia (Brazil) and Minist\'{e}rio de Ciencia, Tecnolog\'{i}a e Innovaci\'{o}n Productiva  (Argentina).
This research has made use of the NASA/IPAC Extragalactic Database (NED) which is operated by the Jet Propulsion Laboratory, California Institute of Technology, under contract with the National Aeronautics and Space Administration.

\begin{deluxetable}{lcccccccccc}  
\tablewidth{0pt}
\tabletypesize{\scriptsize}
\tablecolumns{11}
\tablecaption{Galaxy Properties}
\tablehead{   
  \colhead{Galaxy}	&
  \colhead{R.A.}	&
  \colhead{Dec}	&
  \colhead{Morph.}	&
  \colhead{Distance}	&
  \colhead{D$_{25}$}		&
  \colhead{}	&
  \colhead{}	&
  \colhead{}	&
  \colhead{}	&
  \colhead{Alternate}\\

  \colhead{}			&
  \colhead{(J2000)}			&
  \colhead{(J2000)}			&
  \colhead{type}	&
  \colhead{(Mpc)}		&
  \colhead{(arcmin)}			&
  \colhead{M$_B$}		&
  \colhead{(B-V)$_\circ$}	&
  \colhead{12+log(O/H)}	&
  \colhead{log(N/O)}		&
  \colhead{Name}
}
\startdata
UGC 4305     & 08 19 05.0 & +70 43 12 & Im  	& 3.39  & 7.9 & $-$16.72 & 0.41 & 7.92$\,\pm\,0.10$ & $-$1.52$\,\pm\,0.11$  & Ho II\\
UGC 4459     & 08 34 07.2 & +66 10 54 & Ir	& 3.56  & 1.6 & $-$13.37 & 0.35 & 7.82$\,\pm\,0.09$ & $-$1.32$\,\pm\,0.17$  & DDO 53 \\
UGC 5139     & 09 40 32.3 & +71 10 56 & Ir  	& 3.84  & 3.6 & $-$14.49 & 0.37 & 8.00$\,\pm\,0.10$ & $-$1.39$\,\pm\,0.11$  & Ho I\\
KDG 61     & 09 57 03.1 & +68 35 31 & Sph?  & 3.60  & 2.3 & $-$12.85 & 0.53 & 8.35$\,\pm\,0.05$ & $-$0.96$\,\pm\,0.08$ &\nodata \\
UGC 5336   & 09 57 32.0 & +69 02 45 & Ir  	& 3.70  & 2.6 & $-$13.68 & 0.32 & 8.65$\,\pm\,0.25$ & $-$1.16$\,\pm\,0.13$  & Ho IX\\
Garland    & 10 03 42.0 & +68 41 36 & Im  	& 3.82  & 5.4 & $-$12.40 & \ldots & 8.41$\,\pm\,0.05$ & $-$0.94$\,\pm\,0.15$ &\nodata\\
UGC 5666   & 10 28 23.5 & +68 24 44 & Sm  	& 4.02  & 11.4 & $-$17.46 & 0.47 & 7.93$\,\pm\,0.05$ & $-$1.45$\,\pm\,0.08$  & IC 2574\\
UGC 5692  & 10 30 35.0 & +70 37 07 & Im	& 4.00  & 3.2 & $-$14.63 & 0.72 & 7.95$\,\pm\,0.20$ & $-$1.02$\,\pm\,0.04$  & DDO 82\\
UGC 5918   & 10 49 36.5 & +65 31 50 & Ir  	& 7.40  & 2.4 & $-$14.42 & 0.49 & 7.84$\,\pm\,0.04$ & $-$1.23$\,\pm\,0.07$   & DDO 87\\
UGC 8201  & 13 06 24.8 & +67 42 25 & Im	& 4.57  & 3.2 & $-$15.09 & 0.34 & 7.80$\,\pm\,0.20$ &  \ldots  & DDO 165\\
\enddata
\label{t:param}
\tablecomments{Units of right ascension are hours, minutes, and seconds, and units of declination are degrees, arcminutes, and arcseconds.  Distances, optical sizes, and morphological types are from Karachentsev et al. (2002).  Colors and magnitudes are from Karachentsev et al.\ (2004).  Global oxygen and nitrogen abundances are from the present study.}
\end{deluxetable}

\begin{deluxetable}{ccccc|lcccccl}  
\tablewidth{0pt}
\tabletypesize{\scriptsize}
\tablecolumns{10}
\tablecaption{Slit Positions}
\tablehead{   
 \colhead{Slit}		&
  \colhead{R.A.}	&
  \colhead{Dec}		&
  \colhead{Source\tablenotemark{1}}	&
  \colhead{Redshift\tablenotemark{2}}	&
    \colhead{Slit}	&
  \colhead{R.A.}	&
  \colhead{Dec}		&
  \colhead{Source\tablenotemark{1}}	&
  \colhead{Redshift\tablenotemark{2}}	\\
  
  \colhead{}			&
  \colhead{(J2000)}		&
  \colhead{(J2000)}		&
  \colhead{Type}		&
  \colhead{(z)}			&
    \colhead{}			&
  \colhead{(J2000)}		&
  \colhead{(J2000)}		&
  \colhead{Type}		&
  \colhead{(z)}		
}
\startdata
UGC 4305-2 & 8 19 35.43 & 70 44 23.5 & HII & & UGC 4305-8 & 8 19 09.96 & 70 43 17.9 & HII & \\
UGC 4305-3 & 8 19 30.20 & 70 42 40.5 & HII & & UGC 4305-9 & 8 19 11.90 & 70 43 10.8 & HII &\\
UGC 4305-4 & 8 19 28.35 & 70 42 52.0 & HII & & UGC 4305-11 & 8 19 28.88 & 70 42 19.5  & HII & \\
UGC 4305-5 & 8 19 17.05 & 70 43 41.9 & HII & &UGC 4305-13 & 8 19 27.17 & 70 46 57.1 & HII &  \\
UGC 4305-7 & 8 19 12.54 & 70 43 04.2 & HII & & &&&& \\
\multicolumn{2}{l}{Pos. Angle = 0.15} & \multicolumn{3}{l}{Parallactic Angle = -48.55} & \multicolumn{2}{l}{May 1: 4x 300s} & \multicolumn{2}{l}{Airmass = 1.81} \\
\hline
UGC 4459-2 & 8 34 18.26 & 66 11 24.2 & BG & 0.212&UGC 4459-5 & 8 34 12.44 & 66 10 15.8 & HII &\\
UGC 4459-3 & 8 34 07.23 & 66 10 53.2 & HII & &UGC 4459-6& 8 34 11.26 & 66 10 22.7 & HII & \\
UGC 4459-4 & 8 34 03.66 & 66 10 38.1 &HII & &UGC 4459-7 & 8 34 04.30 & 66 09 05.4 & BG & 0.112\\
\multicolumn{2}{l}{Pos. Angle = 0.05}&\multicolumn{3}{l}{Parallactic Angle = -61.7}&\multicolumn{2}{l}{May 20 \& 21:  4 x 900s}& \multicolumn{2}{l}{Airmass = 1.87} \\
\hline
UGC 5139-1 & 9 40 35.16 & 71 09 54.3 & HII & &UGC 5139-5 & 9 40 44.31 & 71 10 54.5 & HII & \\
UGC 5139-3 & 9 40 36.71 & 71 09 55.6 & HII & & \\
\multicolumn{2}{l}{Pos. Angle = 217.05}&\multicolumn{3}{l}{Parallactic Angle = 310.65}&\multicolumn{2}{l}{May 27 \& 28: 4x 900s}& \multicolumn{2}{l}{Airmass = 1.86} \\
\hline
KDG 61-8 & 9 56 46.25 & 68 35 10.7 & BG & 0.330 & KDG 61-9 & 9 57 07.64 & 68 35 54.2 &  HII &  \\
\multicolumn{2}{l}{Pos. Angle = 271.11}&\multicolumn{3}{l}{Parallactic Angle = 305.37}&\multicolumn{2}{l}{May 28: 4x 900s}& \multicolumn{2}{l}{Airmass = 1.83} \\
\hline
UGC 5336-3& 9 57 26.36 & 69 03 08.8 &  HII & &UGC 5336-10 & 9 57 54.18 & 69 03 48.6 &  SNR &\\
UGC 5336-6& 9 57 33.64 & 69 04 55.1 & BG& 0.175 & UGC 5336-11 & 9 57 52.54 & 69 03 47.1 & SNR & \\
UGC 5336-9& 9 57 47.40 & 69 05 40.4 & BG & 0.294& UGC 5336-12 & 9 57 50.73 & 69 02 22.0 &  HII &\\
\multicolumn{2}{l}{Pos. Angle = 271.07}&\multicolumn{3}{l}{Parallactic Angle = 312.20}&\multicolumn{2}{l}{May 30: 4x 900s}& \multicolumn{2}{l}{Airmass = 1.76} \\
\hline
Garland -1& 10 03 46.27 & 68 41 26.1 & HII &  & Garland -8& 10 04 28.49 & 68 42 38.4 & HII & \\
Garland -3 & 10 04 01.83 & 68 41 24.4 & HII &  & Garland -9& 10 04 26.02 & 68 42 30.2 & HII &  \\
Garland -4& 10 03 57.76 & 68 41 28.2 & HII & &Garland -10& 10 04 27.21 & 68 42 33.4 & H & \\
Garland -5& 10 03 58.75 & 68 40 42.2 & HII &  &Garland -11 & 10 04 14.65 & 68 42 38.3 & H & \\
Garland -6 & 10 04 07.53 & 68 41 05.1 & HII &  & \\
\multicolumn{2}{l}{Pos. Angle = 281.07}&\multicolumn{3}{l}{Parallactic Angle = 310.52}&\multicolumn{2}{l}{May 21: 4x 900s}& \multicolumn{2}{l}{Airmass = 1.77} \\
\hline
UGC 5666-1 & 10 28 48.88 & 68 28 34.4 & HII &  &UGC 5666-9 & 10 28 46.19 & 68 28 25.7 & HII &  \\
UGC 5666-2 & 10 28 50.13 & 68 28 22.2 & HII & & UGC 5666-10 & 10 28 39.34 & 68 28 29.7 & HII & \\
UGC 5666-3 & 10 28 44.14 & 68 28 31.6 & HII & &UGC 5666-11  & 10 28 38.53 & 68 28 12.9 & HII & \\
UGC 5666-5 & 10 28 37.30 & 68 27 56.4 & HII & &UGC 5666-13  & 10 28 58.93 & 68 28 28.2 & HII &  \\
UGC 5666-6 & 10 28 30.54 & 68 28 08.6 & HII & &UGC 5666-14  & 10 28 55.52 & 68 27 55.1 & HII &  \\
UGC 5666-8 & 10 28 49.59 & 68 28 27.9 & HII & & \\
\multicolumn{2}{l}{Pos. Angle = 310.03}&\multicolumn{3}{l}{Parallactic Angle = 307.82}&\multicolumn{2}{l}{May 20: 4x 180s}& \multicolumn{2}{l}{Airmass = 1.77} \\
\hline
UGC 5692-3 & 10 30 35.10 &  70 36 59.6 & HII & & UGC 5692-7& 10 30 36.55 &  70 37 24.5 & H & \\
UGC 5692-4 & 10 30 35.15 &  70 36 51.3 & HII & &UGC 5692-8 & 10 30 33.31 &  70 37 16.7 & H &  \\
UGC 5692-6 & 10 30 33.59 &  70 37 05.9 & HII & & \\
\multicolumn{2}{l}{Pos. Angle = 180.17}&\multicolumn{3}{l}{Parallactic Angle = 149.5}&\multicolumn{2}{l}{May 1: 4x 900s}& \multicolumn{2}{l}{Airmass = 1.67} \\
\hline
UGC 5918-3 & 10 49 43.06 & 65 31 40.5 & HII & &UGC 5918-4 & 10 49 29.77 & 65 31 59.5 & HII & \\
\multicolumn{2}{l}{Pos. Angle = 0.15}&\multicolumn{3}{l}{Parallactic Angle = -60.75}&\multicolumn{2}{l}{May 1:  4 x 900s}& \multicolumn{2}{l}{Airmass = 1.83} \\
\hline
UGC 8201-4 & 13 06 38.00 & 67 42 12.2 & HII & &UGC 8201-15& 13 06 09.66 & 67 43 58.9 & BG & 0.213\\
UGC 8201-8 & 13 06 18.12 & 67 44 16.4 & BG & 0.219 &UGC 8201-11& 13 06 29.96 & 67 42 16.5 & H & \\
UGC 8201-9 & 13 06 24.52 & 67 43 41.2 & HII & & \\
\multicolumn{2}{l}{Pos. Angle = 90.04}&\multicolumn{3}{l}{Parallactic Angle = 333.18}&\multicolumn{2}{l}{May 1: 4x 900s}& \multicolumn{2}{l}{Airmass = 1.56} 
\enddata
\label{t:positions}
\tablenotetext{1}{HII = HII Region, BG=Background Galaxy, SNR=Supernova Remnant, H=Hydrogen Emission Region}
\tablenotetext{2}{Redshifts are only reported for background galaxies.}
\tablecomments{Positions of slit centers for GMOS-N target fields. Units of right ascension are hours, minutes, and seconds, and units of declination are degrees, arcminutes, and arcseconds.  }
\end{deluxetable}

\begin{landscape}
\voffset=0.55in
\begin{deluxetable}{lcccccccccccc}  
\thispagestyle{empty}
\tabletypesize{\scriptsize}
\tablewidth{0pt}
\tablecolumns{12}
\tablecaption{Blue Spectra: Line Intensities Relative to H$\beta$ (= 1)}
\tablehead{   
  \colhead{Slit}	&
  \colhead{[O {\sc ii}]}	&
  \colhead{[Ne {\sc iii}]}	&
  \colhead{H$\gamma$}	&
  \colhead{[O {\sc iii}]}	&
  \colhead{He I}	&
  \colhead{He II}	&
  \colhead{[O {\sc iii}]}	&
  \colhead{[O {\sc iii}]}	&
  \colhead{c$_{H\beta}$}	&
  \colhead{EW(H$\beta$)}	\\

  \colhead{}					&
  \colhead{3727+3729}				&
  \colhead{3869}		&
  \colhead{4340}		&
  \colhead{4363}	&
  \colhead{4471}		&
  \colhead{4686}		&
  \colhead{4959}	&
  \colhead{5007}		&
  \colhead{}	&
  \colhead{$\rm{\AA}$}	
}
\startdata
UGC 4305-2  	&	 0.699$\pm$0.045 	&	\nodata	&	0.479$\pm$0.024	&	0.050$\pm$0.010	&	\nodata	&	\nodata	&	1.327$\pm$0.054	&	3.975$\pm$0.158	&	 0.17$\pm$0.06 	&	 86 \\
UGC 4305-3  	&	 0.776$\pm$0.041 	&	 0.256$\pm$0.014 	&	0.438$\pm$0.019	&	0.070$\pm$0.004	&	\nodata	&	\nodata	&	1.409$\pm$0.053	&	4.212$\pm$0.160	&	 0.06$\pm$0.06 	&	 34 \\
UGC 4305-4  	&	 1.852$\pm$0.102 	&	\nodata	&	0.481$\pm$0.022	&	0.041$\pm$0.007	&	\nodata	&	\nodata	&	0.919$\pm$0.036	&	2.768$\pm$0.108	&	 0.12$\pm$0.06 	&	 14 \\
UGC 4305-5  	&	 2.013$\pm$0.111 	&	\nodata	&	0.597$\pm$0.027	&	0.051$\pm$0.007	&	\nodata	&	\nodata	&	1.109$\pm$0.044	&	3.475$\pm$0.135	&	 0.03$\pm$0.06 	&	 23 \\
UGC 4305-7  	&	 1.638$\pm$0.086 	&	 0.255$\pm$0.014 	&	0.446$\pm$0.019	&	0.053$\pm$0.005	&	0.040$\pm$0.005	&	\nodata	&	1.212$\pm$0.046	&	3.648$\pm$0.139	&	 0.06$\pm$0.06 	&	 32 \\
UGC 4305-8  	&	 2.938$\pm$0.178 	&	\nodata	&	0.527$\pm$0.032	&	\nodata	&	\nodata	&	\nodata	&	0.614$\pm$0.032	&	1.665$\pm$0.074	&	 0.10$\pm$0.06 	&	 9.2\\
UGC 4305-9  	&	 2.329$\pm$0.130 	&	\nodata	&	0.477$\pm$0.024	&	0.060$\pm$0.011	&	\nodata	&	\nodata	&	1.120$\pm$0.046	&	3.341$\pm$0.133	&	 0.09$\pm$0.06 	&	 13 \\
UGC 4305-11 	&	 1.758$\pm$0.092 	&	 0.172$\pm$0.010 	&	0.450$\pm$0.019	&	0.066$\pm$0.009	&	\nodata	&	0.165$\pm$0.007	&	1.044$\pm$0.040	&	3.124$\pm$0.118	&	 0.04$\pm$0.06 	&	 51 \\
UGC 4305-13 	&	 0.768$\pm$0.041 	&	 0.164$\pm$0.009 	&	0.479$\pm$0.020	&	0.050$\pm$0.004	&	0.065$\pm$0.004	&	\nodata	&	1.143$\pm$0.043	&	3.407$\pm$0.129	&	 0.02$\pm$0.06 	&	 50 \\

UGC 4459-3 	&	 0.969$\pm$0.085 	&	 0.179$\pm$0.015 	&	0.445$\pm$0.035	&	0.051$\pm$0.004	&	0.038$\pm$0.003	&	\nodata	&	1.092$\pm$0.081	&	3.334$\pm$0.248	&	 0.09$\pm$0.08 	&	 162\\
UGC 4459-4 	&	 0.911$\pm$0.081 	&	\nodata	&	0.418$\pm$0.033	&	0.019$\pm$0.003	&	0.019$\pm$0.003	&	\nodata	&	0.579$\pm$0.043	&	1.741$\pm$0.130	&	 0.08$\pm$0.08 	&	  48\\
UGC 4459-5 	&	 0.278$\pm$0.025 	&	 0.240$\pm$0.021 	&	0.415$\pm$0.032	&	0.084$\pm$0.007	&	0.038$\pm$0.004	&	\nodata	&	1.482$\pm$0.110	&	4.449$\pm$0.331	&	 0.19$\pm$0.08 	&	  73\\
UGC 4459-6 	&	 1.006$\pm$0.090 	&	 0.150$\pm$0.014 	&	0.416$\pm$0.033	&	0.030$\pm$0.005	&	0.038$\pm$0.005	&	\nodata	&	1.011$\pm$0.075	&	3.012$\pm$0.225	&	 0.10$\pm$0.08 	&	  55\\

UGC 5139-1 	&	 0.742$\pm$0.044 	&	 0.137$\pm$0.010 	&	0.415$\pm$0.020	&	0.036$\pm$0.006	&	\nodata	&	\nodata	&	1.192$\pm$0.048	&	4.114$\pm$0.166	&	 0.41$\pm$0.06 	&	 102  \\
UGC 5139-3 	&	 1.550$\pm$0.142 	&	 0.271$\pm$0.027 	&	0.471$\pm$0.024	&	0.057$\pm$0.009	&	\nodata	&	\nodata	&	0.894$\pm$0.038	&	2.780$\pm$0.115	&	 0.40$\pm$0.08 	&	  26  \\
UGC 5139-5 	&	 1.531$\pm$0.086 	&	 0.139$\pm$0.010 	&	0.420$\pm$0.019	&	0.032$\pm$0.005	&	0.025$\pm$0.004	&	\nodata	&	1.116$\pm$0.045	&	3.386$\pm$0.136	&	 0.52$\pm$0.06 	&	 194  \\

KDG 61-9 	&	 1.287$\pm$0.067 	&	 0.538$\pm$0.028 	&	0.465$\pm$0.021	&	0.065$\pm$0.008	&	\nodata	&	0.112$\pm$0.009	&	2.422$\pm$0.091	&	7.295$\pm$0.275	&	 0.06$\pm$0.05 	&	 28\\

UGC 5336-3  	&	 2.632$\pm$0.160 	&	\nodata	&	0.567$\pm$0.031	&	\nodata	&	\nodata	&	\nodata	&	0.079$\pm$0.013	&	0.334$\pm$0.019	&	 0.14$\pm$0.06 	&	  65 \\
UGC 5336-10 	&	 4.374$\pm$0.248 	&	\nodata	&	0.546$\pm$0.030	&	\nodata	&	\nodata	&	\nodata	&	0.477$\pm$0.023	&	1.560$\pm$0.064	&	 0.52$\pm$0.05 	&	 1095\\
UGC 5336-11 	&	 6.234$\pm$0.349 	&	\nodata	&	0.448$\pm$0.025	&	\nodata	&	\nodata	&	\nodata	&	0.517$\pm$0.024	&	1.784$\pm$0.073	&	 0.34$\pm$0.06 	&	 508 \\
UGC 5336-12 	&	 3.456$\pm$0.281 	&	\nodata	&	0.767$\pm$0.066	&	\nodata	&	\nodata	&	\nodata	&	0.170$\pm$0.039	&	0.315$\pm$0.042	&	 0.27$\pm$0.08 	&	  16 \\

Garland-1 	&	 2.459$\pm$0.133 	&	\nodata	&	0.438$\pm$0.021	&	\nodata	&	\nodata	&	\nodata	&	0.260$\pm$0.012	&	0.844$\pm$0.033	&	 0.34$\pm$0.05 	&	   141\\
Garland-2 	&	 9.284$\pm$0.553 	&	\nodata	&	0.884$\pm$0.056	&	\nodata	&	\nodata	&	\nodata	&	0.367$\pm$0.029	&	1.062$\pm$0.051	&	 0.82$\pm$0.06 	&	    71\\
Garland-3 	&	 2.313$\pm$0.095 	&	\nodata	&	0.444$\pm$0.015	&	\nodata	&	0.029$\pm$0.004	&	\nodata	&	0.883$\pm$0.025	&	2.777$\pm$0.078	&	 0.40$\pm$0.05 	&	   140\\
Garland-4 	&	 2.434$\pm$0.117 	&	\nodata	&	0.463$\pm$0.020	&	\nodata	&	\nodata	&	\nodata	&	0.703$\pm$0.023	&	2.070$\pm$0.063	&	 0.66$\pm$0.05 	&	   105\\
Garland-5 	&	 2.067$\pm$0.109 	&	 0.801$\pm$0.042 	&	0.464$\pm$0.021	&	0.069$\pm$0.007	&	\nodata	&	\nodata	&	2.548$\pm$0.096	&	7.573$\pm$0.285	&	 0.21$\pm$0.05 	&	  1632\\
Garland-6 	&	 2.907$\pm$0.125 	&	\nodata	&	0.475$\pm$0.018	&	\nodata	&	\nodata	&	\nodata	&	0.058$\pm$0.006	&	0.239$\pm$0.009	&	 0.31$\pm$0.05 	&	    84\\
Garland-8 	&	 2.880$\pm$0.122 	&	\nodata	&	0.465$\pm$0.018	&	\nodata	&	\nodata	&	\nodata	&	0.620$\pm$0.020	&	1.861$\pm$0.055	&	 0.22$\pm$0.05 	&	    99\\
Garland-9 	&	 2.457$\pm$0.124 	&	\nodata	&	0.431$\pm$0.024	&	\nodata	&	\nodata	&	\nodata	&	0.149$\pm$0.016	&	0.532$\pm$0.023	&	 0.25$\pm$0.05 	&	    45\\

UGC 5666-1  	&	 1.150$\pm$0.102 	&	 0.346$\pm$0.030 	&	0.454$\pm$0.035	&	0.072$\pm$0.006	&	0.029$\pm$0.003	&	\nodata	&	1.476$\pm$0.110	&	4.409$\pm$0.328	&	 0.12$\pm$0.08 	&	 214\\
UGC 5666-2  	&	 1.885$\pm$0.169 	&	 0.209$\pm$0.019 	&	0.432$\pm$0.034	&	0.032$\pm$0.004	&	0.030$\pm$0.004	&	\nodata	&	0.903$\pm$0.067	&	2.690$\pm$0.201	&	 0.25$\pm$0.08 	&	 129\\
UGC 5666-3  	&	 1.066$\pm$0.095 	&	 0.441$\pm$0.038 	&	0.461$\pm$0.036	&	0.080$\pm$0.007	&	0.032$\pm$0.003	&	\nodata	&	1.565$\pm$0.116	&	4.689$\pm$0.349	&	 0.11$\pm$0.08 	&	 388\\
UGC 5666-5  	&	 1.450$\pm$0.130 	&	 0.369$\pm$0.033 	&	0.467$\pm$0.037	&	0.066$\pm$0.006	&	\nodata	&	\nodata	&	1.516$\pm$0.113	&	4.574$\pm$0.341	&	 0.05$\pm$0.08 	&	 151\\
UGC 5666-6  	&	 1.505$\pm$0.135 	&	 0.325$\pm$0.029 	&	0.493$\pm$0.039	&	0.053$\pm$0.005	&	0.028$\pm$0.004	&	\nodata	&	1.362$\pm$0.102	&	4.122$\pm$0.307	&	 0.07$\pm$0.08 	&	 154\\
UGC 5666-8  	&	 1.259$\pm$0.113 	&	 0.323$\pm$0.029 	&	0.455$\pm$0.036	&	0.056$\pm$0.006	&	0.033$\pm$0.005	&	\nodata	&	1.383$\pm$0.103	&	4.148$\pm$0.309	&	 0.09$\pm$0.08 	&	 160\\
UGC 5666-9  	&	 1.408$\pm$0.132 	&	 0.394$\pm$0.039 	&	0.481$\pm$0.040	&	0.080$\pm$0.012	&	\nodata	&	\nodata	&	1.162$\pm$0.088	&	3.463$\pm$0.262	&	 0.07$\pm$0.08 	&	  89\\
UGC 5666-10 	&	 1.383$\pm$0.130 	&	 0.424$\pm$0.041 	&	0.470$\pm$0.039	&	0.052$\pm$0.009	&	\nodata	&	\nodata	&	1.250$\pm$0.095	&	3.704$\pm$0.279	&	 0.19$\pm$0.08 	&	  65\\
UGC 5666-11 	&	 2.133$\pm$0.198 	&	 0.207$\pm$0.023 	&	0.413$\pm$0.035	&	\nodata	&	\nodata	&	\nodata	&	0.754$\pm$0.058	&	2.640$\pm$0.200	&	 0.08$\pm$0.08 	&	  98\\
UGC 5666-13 	&	 1.246$\pm$0.112 	&	 0.290$\pm$0.026 	&	0.452$\pm$0.036	&	0.063$\pm$0.007	&	\nodata	&	\nodata	&	1.253$\pm$0.093	&	3.710$\pm$0.277	&	 0.15$\pm$0.08 	&	 110\\
UGC 5666-14 	&	 1.795$\pm$0.178 	&	 0.516$\pm$0.056 	&	0.460$\pm$0.042	&	\nodata	&	\nodata	&	\nodata	&	1.303$\pm$0.101	&	3.917$\pm$0.301	&	 0.41$\pm$0.08 	&	  48\\

UGC 5692-3 	&	 3.278$\pm$0.145 	&	\nodata	&	0.552$\pm$0.023	&	\nodata	&	\nodata	&	\nodata	&	0.356$\pm$0.015	&	1.024$\pm$0.030	&	 0.21$\pm$0.04 	&	 16 \\
UGC 5692-4 	&	 2.397$\pm$0.113 	&	\nodata	&	0.479$\pm$0.020	&	\nodata	&	\nodata	&	\nodata	&	0.732$\pm$0.022	&	2.284$\pm$0.060	&	 0.30$\pm$0.05 	&	 29 \\
UGC 5692-6 	&	 3.319$\pm$0.147 	&	\nodata	&	0.677$\pm$0.026	&	\nodata	&	\nodata	&	\nodata	&	0.600$\pm$0.020	&	1.419$\pm$0.039	&	 0.45$\pm$0.05 	&	 14 \\

UGC 5918-3 	&	 1.207$\pm$0.046 	&	\nodata	&	0.458$\pm$0.013	&	0.041$\pm$0.004	&	0.040$\pm$0.004	&	\nodata	&	1.046$\pm$0.023	&	3.018$\pm$0.067	&	 0.02$\pm$0.04 	&	 177 \\
UGC 5918-4 	&	 1.705$\pm$0.079 	&	\nodata	&	0.437$\pm$0.018	&	\nodata	&	\nodata	&	\nodata	&	0.757$\pm$0.022	&	2.205$\pm$0.057	&	 0.02$\pm$0.05 	&	 71 \\

UGC 8201-4 	&	 2.254$\pm$0.391 	&	 0.146$\pm$0.117 	&	0.423$\pm$0.126	&	\nodata	&	\nodata	&	\nodata	&	0.519$\pm$0.127	&	1.639$\pm$0.241	&	 0.11$\pm$0.17 	&	 33\\
UGC 8201-9 	&	 2.565$\pm$0.548 	&	\nodata	&	0.434$\pm$0.166	&	\nodata	&	\nodata	&	\nodata	&	0.301$\pm$0.150	&	1.014$\pm$0.217	&	 0.10$\pm$0.21 	&	44\thispagestyle{empty}
\enddata
\label{t:lineratiob}
\end{deluxetable}
\voffset=0.55in
\end{landscape}

\begin{landscape}
\voffset=0.55in
\begin{deluxetable}{lcccccccccccc}  
\thispagestyle{empty}
\tabletypesize{\scriptsize}
\tablewidth{0pt}
\tablecolumns{12}
\tablecaption{Red spectra: Line Intensities Relative to H$\beta$ (= 1)}
\tablehead{   
  \colhead{Slit}	&
  \colhead{[O {\sc i}]}	&
  \colhead{[S {\sc iii}]}	&
  \colhead{[O {\sc i}]}	&
  \colhead{[N~\ii]}	&
  \colhead{H$\alpha$}	&
  \colhead{[N~\ii]}	&
  \colhead{He {\sc i}}	&
  \colhead{[S {\sc ii}]}	&
  \colhead{[S {\sc ii}]}	&
  \colhead{He {\sc i}}	&
  \colhead{[Ar {\sc iii}]}\\

  \colhead{}		&
  \colhead{6300}	&
  \colhead{6311}	&
  \colhead{6364}	&
  \colhead{6548}	&
  \colhead{6563}	&
  \colhead{6584}	&
  \colhead{6678}	&
  \colhead{6716}	&
  \colhead{6731}	&
  \colhead{7065}	&
  \colhead{7136}	
}
\startdata
UGC 4305-2  	&	\nodata	&	 0.017$\pm$0.004 	&	\nodata	&	0.022$\pm$0.004	&	 2.837$\pm$0.180 	&	0.040$\pm$0.005	&	0.023$\pm$0.004	&	0.089$\pm$0.007	&	0.062$\pm$0.006	&	0.025$\pm$0.004	&	 0.051$\pm$0.005 	\\
UGC 4305-3  	&	 0.075$\pm$0.005 	&	 0.018$\pm$0.002 	&	 0.019$\pm$0.002 	&	0.012$\pm$0.002	&	 2.707$\pm$0.165 	&	0.028$\pm$0.003	&	0.026$\pm$0.003	&	0.079$\pm$0.005	&	0.058$\pm$0.004	&	0.022$\pm$0.003	&	 0.048$\pm$0.004 	\\
UGC 4305-4  	&	\nodata	&	\nodata	&	\nodata	&	0.036$\pm$0.005	&	 2.831$\pm$0.176 	&	0.059$\pm$0.006	&	\nodata	&	0.166$\pm$0.012	&	0.119$\pm$0.009	&	\nodata	&	 0.036$\pm$0.005 	\\
UGC 4305-5  	&	 0.256$\pm$0.015 	&	 0.020$\pm$0.004 	&	 0.053$\pm$0.005 	&	0.023$\pm$0.004	&	 3.103$\pm$0.193 	&	0.056$\pm$0.005	&	0.031$\pm$0.004	&	0.125$\pm$0.009	&	0.096$\pm$0.007	&	\nodata	&	 0.062$\pm$0.006 	\\
UGC 4305-7  	&	\nodata	&	 0.016$\pm$0.002 	&	\nodata	&	0.030$\pm$0.002	&	 2.741$\pm$0.167 	&	0.083$\pm$0.005	&	0.037$\pm$0.003	&	0.173$\pm$0.011	&	0.128$\pm$0.008	&	0.020$\pm$0.002	&	 0.062$\pm$0.005 	\\
UGC 4305-8  	&	\nodata	&	 0.016$\pm$0.007 	&	\nodata	&	0.050$\pm$0.008	&	 2.932$\pm$0.197 	&	0.134$\pm$0.012	&	0.030$\pm$0.007	&	0.315$\pm$0.023	&	0.254$\pm$0.019	&	\nodata	&	 0.034$\pm$0.007 	\\
UGC 4305-9  	&	\nodata	&	 0.017$\pm$0.004 	&	\nodata	&	0.039$\pm$0.004	&	 2.804$\pm$0.178 	&	0.079$\pm$0.006	&	0.030$\pm$0.004	&	0.165$\pm$0.011	&	0.119$\pm$0.009	&	\nodata	&	 0.048$\pm$0.005 	\\
UGC 4305-11 	&	 0.094$\pm$0.006 	&	 0.013$\pm$0.002 	&	 0.023$\pm$0.002 	&	0.034$\pm$0.003	&	 2.722$\pm$0.166	&	0.079$\pm$0.005	&	0.026$\pm$0.002	&	0.194$\pm$0.012	&	0.142$\pm$0.009	&	0.012$\pm$0.002	&	 0.032$\pm$0.003 	\\
UGC 4305-13 	&	\nodata	&	 0.010$\pm$0.002 	&	\nodata	&	0.011$\pm$0.002	&	 2.832$\pm$0.172 	&	0.028$\pm$0.002	&	0.032$\pm$0.003	&	0.061$\pm$0.004	&	0.041$\pm$0.003	&	0.019$\pm$0.002	&	 0.038$\pm$0.003 	\\

UGC 4459-3 	&	 0.049$\pm$0.004 	&	 0.014$\pm$0.001 	&	 0.022$\pm$0.003 	&	0.016$\pm$0.002	&	 2.767$\pm$0.224 	&	0.057$\pm$0.005	&	0.027$\pm$0.002	&	0.146$\pm$0.012	&	0.105$\pm$0.009	&	0.018$\pm$0.002	&	 0.048$\pm$0.004 	\\
UGC 4459-4 	&	\nodata	&	 0.012$\pm$0.003 	&	\nodata	&	0.025$\pm$0.003	&	 2.646$\pm$0.215 	&	0.081$\pm$0.007	&	0.027$\pm$0.003	&	0.145$\pm$0.012	&	0.110$\pm$0.010	&	0.012$\pm$0.003	&	 0.044$\pm$0.005 	\\
UGC 4459-5 	&	\nodata	&	 0.012$\pm$0.001 	&	 0.065$\pm$0.005 	&	0.016$\pm$0.002	&	 2.704$\pm$0.220 	&	0.029$\pm$0.003	&	0.031$\pm$0.003	&	0.033$\pm$0.003	&	0.034$\pm$0.003	&	0.053$\pm$0.005	&	 0.055$\pm$0.005 	\\
UGC 4459-6 	&	\nodata	&	 0.016$\pm$0.002 	&	\nodata	&	0.013$\pm$0.002	&	 2.644$\pm$0.216 	&	0.056$\pm$0.005	&	0.022$\pm$0.003	&	0.115$\pm$0.010	&	0.085$\pm$0.008	&	0.018$\pm$0.003	&	 0.048$\pm$0.005 	\\

UGC 5139-1 	&	\nodata	&	 0.013$\pm$0.002 	&	\nodata	&	0.018$\pm$0.002	&	 2.689$\pm$0.178 	&	0.063$\pm$0.005	&	0.032$\pm$0.003	&	0.145$\pm$0.010	&	0.103$\pm$0.007	&	0.016$\pm$0.002	&	 0.078$\pm$0.006 	\\
UGC 5139-3 	&	\nodata	&	\nodata	&	\nodata	&	0.016$\pm$0.005	&	 2.759$\pm$0.236 	&	0.088$\pm$0.008	&	0.048$\pm$0.006	&	0.169$\pm$0.013	&	0.120$\pm$0.010	&	\nodata	&	 0.070$\pm$0.008 	\\
UGC 5139-5 	&	\nodata	&	 0.015$\pm$0.002 	&	\nodata	&	0.027$\pm$0.002	&	 2.694$\pm$0.177 	&	0.085$\pm$0.006	&	0.035$\pm$0.003	&	0.176$\pm$0.012	&	0.127$\pm$0.009	&	0.020$\pm$0.002	&	 0.071$\pm$0.005 	\\

KDG 61-9 	&	 0.083$\pm$0.005 	&	 \ldots 	&	 0.047$\pm$0.004 	&	0.073$\pm$0.005	&	 2.834$\pm$0.155 	&	0.221$\pm$0.012	&	0.024$\pm$0.003	&	0.166$\pm$0.010	&	0.107$\pm$0.007	&	0.014$\pm$0.003	&	 0.079$\pm$0.006 	\\

UGC 5336-3  	&	\nodata	&	\nodata	&	 0.177$\pm$0.013 	&	0.179$\pm$0.014	&	 2.995$\pm$0.196 	&	0.473$\pm$0.032	&	\nodata	&	0.375$\pm$0.026	&	0.281$\pm$0.020	&	\nodata	&	0.022$\pm$0.007	\\
UGC 5336-10 	&	\nodata	&	\nodata	&	\nodata	&	0.287$\pm$0.018	&	 2.927$\pm$0.170 	&	0.807$\pm$0.048	&	0.024$\pm$0.007	&	1.073$\pm$0.065	&	0.720$\pm$0.044	&	0.008$\pm$0.006	&	 0.052$\pm$0.01 	\\
UGC 5336-11 	&	\nodata	&	\nodata	&	\nodata	&	0.336$\pm$0.022	&	 2.777$\pm$0.179 	&	0.882$\pm$0.057	&	0.029$\pm$0.005	&	1.186$\pm$0.079	&	0.815$\pm$0.055	&	\nodata	&	 0.046$\pm$0.01 	\\
UGC 5336-12 	&	\nodata	&	\nodata	&	\nodata	&	0.209$\pm$0.021	&	 3.137$\pm$0.266 	&	0.543$\pm$0.048	&	\nodata	&	0.454$\pm$0.041	&	0.325$\pm$0.031	&	\nodata	&	\nodata	\\

Garland-1 	&	\nodata	&	\nodata	&	\nodata	&	0.244$\pm$0.014	&	 2.759$\pm$0.152 	&	0.583$\pm$0.033	&	0.029$\pm$0.004	&	0.319$\pm$0.019	&	0.226$\pm$0.014	&	0.016$\pm$0.004	&	 0.061$\pm$0.005 	\\
Garland-2 	&	 0.531$\pm$0.032 	&	\nodata	&	 0.252$\pm$0.018 	&	0.200$\pm$0.015	&	 3.221$\pm$0.197 	&	0.477$\pm$0.031	&	0.051$\pm$0.009	&	0.334$\pm$0.023	&	0.249$\pm$0.018	&	\nodata	&	 0.049$\pm$0.009 	\\
Garland-3 	&	\nodata	&	\nodata	&	 0.052$\pm$0.003 	&	0.115$\pm$0.006	&	 2.757$\pm$0.125 	&	0.311$\pm$0.014	&	0.031$\pm$0.003	&	0.168$\pm$0.008	&	0.112$\pm$0.006	&	0.014$\pm$0.002	&	 0.070$\pm$0.004 	\\
Garland-4 	&	\nodata	&	\nodata	&	 0.118$\pm$0.007 	&	0.168$\pm$0.009	&	 2.807$\pm$0.134 	&	0.388$\pm$0.019	&	0.046$\pm$0.004	&	0.208$\pm$0.011	&	0.155$\pm$0.009	&	0.016$\pm$0.003	&	 0.080$\pm$0.005 	\\
Garland-5 	&	\nodata	&	\nodata	&	 0.066$\pm$0.005 	&	0.136$\pm$0.008	&	 2.830$\pm$0.155 	&	0.336$\pm$0.019	&	0.045$\pm$0.004	&	0.287$\pm$0.017	&	0.197$\pm$0.012	&	0.021$\pm$0.003	&	 0.117$\pm$0.008 	\\
Garland-6 	&	\nodata	&	\nodata	&	 0.108$\pm$0.006 	&	0.213$\pm$0.010	&	 2.828$\pm$0.131 	&	0.503$\pm$0.024	&	0.026$\pm$0.003	&	0.383$\pm$0.019	&	0.257$\pm$0.013	&	\nodata	&	 0.025$\pm$0.003 	\\
Garland-8 	&	\nodata	&	\nodata	&	\nodata	&	0.183$\pm$0.010	&	 2.809$\pm$0.132 	&	0.452$\pm$0.022	&	0.030$\pm$0.006	&	0.290$\pm$0.015	&	0.196$\pm$0.011	&	0.019$\pm$0.005	&	 0.058$\pm$0.006 	\\
Garland-9 	&	\nodata	&	\nodata	&	\nodata	&	0.138$\pm$0.009	&	 2.772$\pm$0.143 	&	0.420$\pm$0.023	&	0.050$\pm$0.006	&	0.331$\pm$0.019	&	0.220$\pm$0.013	&	\nodata	&	 0.040$\pm$0.006 	\\

UGC 5666-1  	&	 0.043$\pm$0.003 	&	 0.017$\pm$0.001 	&	 0.015$\pm$0.001 	&	0.020$\pm$0.002	&	 2.774$\pm$0.233 	&	0.046$\pm$0.004	&	0.031$\pm$0.003	&	0.088$\pm$0.008	&	0.063$\pm$0.006	&	0.019$\pm$0.002	&	 0.054$\pm$0.005 	\\
UGC 5666-2  	&	 0.066$\pm$0.005 	&	\nodata	&	 0.024$\pm$0.002 	&	0.035$\pm$0.003	&	 2.762$\pm$0.233 	&	0.094$\pm$0.008	&	0.030$\pm$0.003	&	0.179$\pm$0.016	&	0.119$\pm$0.010	&	0.018$\pm$0.002	&	 0.052$\pm$0.005 	\\
UGC 5666-3  	&	 0.030$\pm$0.002 	&	 0.020$\pm$0.002 	&	 0.014$\pm$0.001 	&	0.022$\pm$0.002	&	 2.782$\pm$0.234 	&	0.058$\pm$0.005	&	0.030$\pm$0.003	&	0.124$\pm$0.011	&	0.085$\pm$0.007	&	0.018$\pm$0.002	&	 0.054$\pm$0.005 	\\
UGC 5666-5  	&	 0.031$\pm$0.003 	&	 0.024$\pm$0.002 	&	 0.014$\pm$0.002 	&	0.021$\pm$0.002	&	 2.804$\pm$0.237 	&	0.056$\pm$0.005	&	0.032$\pm$0.003	&	0.126$\pm$0.011	&	0.088$\pm$0.008	&	0.021$\pm$0.002	&	 0.061$\pm$0.006 	\\
UGC 5666-6  	&	\nodata	&	 0.013$\pm$0.002 	&	 0.007$\pm$0.001 	&	0.023$\pm$0.002	&	 2.850$\pm$0.241 	&	0.057$\pm$0.005	&	0.027$\pm$0.003	&	0.117$\pm$0.010	&	0.085$\pm$0.008	&	0.021$\pm$0.002	&	 0.058$\pm$0.006 	\\
UGC 5666-8  	&	 0.059$\pm$0.005 	&	 0.019$\pm$0.002 	&	 0.022$\pm$0.002 	&	0.023$\pm$0.002	&	 2.789$\pm$0.235 	&	0.067$\pm$0.006	&	0.031$\pm$0.003	&	0.139$\pm$0.012	&	0.094$\pm$0.008	&	0.020$\pm$0.002	&	 0.058$\pm$0.006 	\\
UGC 5666-9  	&	 0.141$\pm$0.012 	&	 0.024$\pm$0.003 	&	 0.077$\pm$0.007 	&	0.035$\pm$0.004	&	 2.790$\pm$0.240 	&	0.093$\pm$0.008	&	0.040$\pm$0.004	&	0.214$\pm$0.019	&	0.145$\pm$0.013	&	0.019$\pm$0.003	&	 0.042$\pm$0.005 	\\
UGC 5666-10 	&	 0.046$\pm$0.004 	&	\nodata	&	 0.036$\pm$0.004 	&	0.014$\pm$0.002	&	 2.811$\pm$0.241 	&	0.052$\pm$0.005	&	0.034$\pm$0.004	&	0.136$\pm$0.012	&	0.093$\pm$0.009	&	0.022$\pm$0.003	&	 0.051$\pm$0.005 	\\
UGC 5666-11 	&	 0.092$\pm$0.008 	&	 0.014$\pm$0.003 	&	 0.036$\pm$0.004 	&	\nodata	&	 2.735$\pm$0.235 	&	0.093$\pm$0.009	&	0.032$\pm$0.004	&	0.198$\pm$0.018	&	0.136$\pm$0.013	&	0.023$\pm$0.004	&	 0.044$\pm$0.005 	\\
UGC 5666-13 	&	\nodata	&	 0.017$\pm$0.002 	&	 0.015$\pm$0.002 	&	0.020$\pm$0.002	&	 2.769$\pm$0.234 	&	0.063$\pm$0.005	&	0.032$\pm$0.003	&	0.148$\pm$0.013	&	0.106$\pm$0.009	&	\nodata	&	 0.050$\pm$0.005 	\\
UGC 5666-14 	&	 0.232$\pm$0.020 	&	\nodata	&	 0.081$\pm$0.008 	&	\nodata	&	 2.806$\pm$0.248 	&	0.068$\pm$0.007	&	0.032$\pm$0.005	&	0.132$\pm$0.013	&	0.091$\pm$0.009	&	0.066$\pm$0.008	&	 0.047$\pm$0.006 	\\

UGC 5692-3 	&	 0.200$\pm$0.011 	&	\nodata	&	 0.090$\pm$0.006 	&	0.129$\pm$0.008	&	 2.954$\pm$0.140 	&	0.449$\pm$0.022	&	0.040$\pm$0.005	&	0.759$\pm$0.038	&	0.550$\pm$0.028	&	0.012$\pm$0.005	&	 0.067$\pm$0.006 	\\
UGC 5692-4 	&	 0.439$\pm$0.020 	&	\nodata	&	\nodata	&	0.103$\pm$0.006	&	 2.833$\pm$0.132 	&	0.301$\pm$0.015	&	0.028$\pm$0.004	&	0.358$\pm$0.018	&	0.262$\pm$0.013	&	0.023$\pm$0.004	&	 0.131$\pm$0.008 	\\
UGC 5692-6 	&	 0.215$\pm$0.011 	&	\nodata	&	 0.061$\pm$0.005 	&	0.164$\pm$0.009	&	 3.169$\pm$0.149 	&	0.428$\pm$0.021	&	\nodata	&	0.822$\pm$0.040	&	0.563$\pm$0.028	&	\nodata	&	 0.079$\pm$0.006 	\\

UGC 5918-3 	&	\nodata	&	 0.017$\pm$0.002 	&	\nodata	&	0.034$\pm$0.002	&	 2.780$\pm$0.119 	&	0.099$\pm$0.005	&	0.026$\pm$0.002	&	0.180$\pm$0.008	&	0.123$\pm$0.006	&	0.018$\pm$0.002	&	 0.068$\pm$0.004 	\\
UGC 5918-4 	&	\nodata	&	\nodata	&	\nodata	&	\nodata	&	 2.737$\pm$0.127 	&	0.067$\pm$0.005	&	\nodata	&	0.184$\pm$0.010	&	0.129$\pm$0.008	&	0.025$\pm$0.004	&	 0.073$\pm$0.006 	\\

UGC 8201-4 	&	 0.290$\pm$0.048 	&	\nodata	&	 0.116$\pm$0.020 	&	\nodata	&	 2.747$\pm$0.479 	&	\nodata	&	0.025$\pm$0.007	&	0.158$\pm$0.029	&	0.122$\pm$0.023	&	\nodata	&	\nodata	\\
UGC 8201-9 	&	\nodata	&	\nodata	&	 0.292$\pm$0.062 	&	\nodata	&	 2.804$\pm$0.614 	&	\nodata	&	\nodata	&	0.113$\pm$0.027	&	0.075$\pm$0.020	&	\nodata	&	\nodata	
\thispagestyle{empty}
\enddata
\label{t:lineratior}
\end{deluxetable}
\voffset=0.55in
\end{landscape}

\voffset=0.0in
\begin{deluxetable}{lcccccccc}  
\tabletypesize{\scriptsize}
\tablewidth{0pt}
\tablecolumns{9}
\tablecaption{H {\sc ii} Region Abundances}
\tablehead{   
  \colhead{Slit}	&
  \colhead{(O/H)$_{McG}$}	&
  \colhead{12+log(O/H)}	&
  \colhead{log(N/O)}		&
  \colhead{log(S/O)}	&
  \colhead{log(Ne/O)}		&
  \colhead{log(Ar/O)}		&
  \colhead{T(O {\sc iii})}	&
  \colhead{S/N$_{4363}$}	
}
\startdata
UGC 4305-2   & 7.70 & 7.92$\pm$0.07 & -1.34$\pm$0.11 & -1.34$\pm$0.19 &    \ldots      & -2.08$\pm$0.13 & 12522$\pm^{1200}_{1100}$ & 3.5\\
UGC 4305-3   & 7.76 & 7.85$\pm$0.03 & -1.53$\pm$0.08 & -1.43$\pm$0.09 & -0.94$\pm$0.08 & -2.15$\pm$0.09 & 14057$\pm^{590}_{560}$ & 9.1\\
UGC 4305-4   & 7.81 & 7.82$\pm$0.06 & -1.57$\pm$0.10 &    \ldots      &   \ldots       & -2.40$\pm$0.13 & 13450$\pm^{1130}_{1170}$ & 2.8\\
UGC 4305-5   & 7.94 & 7.98$\pm$0.04 & -1.67$\pm$0.09 & -1.59$\pm$0.14 &   \ldots       & -2.29$\pm$0.10 & 13432$\pm^{960}_{970}$ & 3.2\\
UGC 4305-7   & 7.89 & 7.89$\pm$0.04 & -1.41$\pm$0.08 & -1.54$\pm$0.10 & -0.87$\pm$0.10 & -2.20$\pm$0.09 & 13350$\pm^{680}_{750}$ & 6.8\\
UGC 4305-8   & 7.93 & 7.93$\pm$0.20 & -1.49$\pm$0.16 & -1.45$\pm$0.26 &   \ldots       & -2.42$\pm$0.21 & 12070$\pm^{1830}_{1770}$ & $<$2.5\\
UGC 4305-9   & 8.01 & 7.84$\pm$0.06 & -1.52$\pm$0.10 & -1.61$\pm$0.15 &   \ldots       & -2.34$\pm$0.13 & 14579$\pm^{1430}_{1450}$ & 3.0\\
UGC 4305-11  & 7.85 & 7.75$\pm$0.05 & -1.40$\pm$0.09 & -1.60$\pm$0.11 & -1.00$\pm$0.13 & -2.49$\pm$0.11 & 15618$\pm^{1390}_{1220}$ & 7.7\\
UGC 4305-13  & 7.66 & 7.80$\pm$0.04 & -1.56$\pm$0.08 & -1.60$\pm$0.12 & -1.03$\pm$0.09 & -2.20$\pm$0.09 & 13298$\pm^{700}_{580}$ & 5.1\\

UGC 4459-3   &   7.69 &7.81$\pm$0.05 &-1.40$\pm$0.09 &-1.49$\pm$0.10& -0.98$\pm$0.12& -2.17$\pm$0.10& 13657$\pm^{860}_{870}  $ & 21\\
UGC 4459-4   &   7.46 &7.72$\pm$0.06 &-1.24$\pm$0.11 &-1.34$\pm$0.16&  \ldots       & -2.08$\pm$0.13& 11862$\pm^{1110}_{910} $ & 3.2\\
UGC 4459-5   &   7.60 &7.83$\pm$0.05 &-1.13$\pm$0.10 &-1.37$\pm$0.12& -0.99$\pm$0.12& -1.91$\pm$0.10& 14882$\pm^{1130}_{1010}$ & 21\\
UGC 4459-6   &   7.67 &7.95$\pm$0.07 &-1.52$\pm$0.11 &-1.37$\pm$0.16& -0.98$\pm$0.17& -2.16$\pm$0.13& 11509$\pm^{1040}_{920} $ & 2.9 \\ 

UGC 5139-1     &   7.71 &8.04$\pm$0.07 &-1.29$\pm$0.09 &-1.37$\pm$0.15& -1.12$\pm$0.15& -1.91$\pm$0.12& 11210$\pm^{740}_{830}  $ & 4.4\\
UGC 5139-5     &   7.82 &8.04$\pm$0.06 &-1.45$\pm$0.09 &  \ldots      & -1.06$\pm$0.14& -2.12$\pm$0.11& 11365$\pm^{750}_{790}  $ & 5.5\\
UGC 5139-3     &   7.78 &7.86$\pm$0.09 &-1.51$\pm$0.14 &  \ldots      & -0.74$\pm$0.23& -2.09$\pm$0.16& 12907$\pm^{1790}_{1560}$ & 2.9 \\

KDG 61-9   &   8.32 &8.35$\pm$0.05 &-0.96$\pm$0.08 &  \ldots      & -0.80$\pm$0.12& -2.18$\pm$0.10& 11117$\pm^{600}_{610}  $ & 6.0\\

UGC 5336-3    &   8.91 & 8.91$\pm$0.20& -1.24$\pm$0.18&  \ldots      &  \ldots       & \ldots &  5550$\pm^{620}_{470}   $ & $<$2.5\\ 
UGC 5336-12   &   8.82 & 8.82$\pm$0.20& -1.31$\pm$0.13&  \ldots      &  \ldots       &  \ldots       &  5500$\pm^{360}_{310 }  $ & $<$2.5 \\
UGC 5336-10\tablenotemark{a}   &   8.44 & 8.44$\pm$0.20& -1.02$\pm$0.14&  \ldots      &  \ldots       & -2.34$\pm$0.18&  8330$\pm^{700}_{590  } $ & $<$2.5 \\
UGC 5336-11\tablenotemark{a}   &   8.43 & 8.43$\pm$0.20& -1.08$\pm$0.14&  \ldots      & \ldots & -2.39$\pm$0.18&  9190$\pm^{980}_{800   }$ & $<$2.5 \\

Garland-5 &   8.32 &8.41$\pm$0.04 &-0.94$\pm$0.08 &  \ldots      & -0.65$\pm$0.11& -2.14$\pm$0.09& 11200$\pm^{510}_{550}  $ & 4.5 \\
Garland-1 &   8.89 &8.86$\pm$0.20 &-1.10$\pm$0.15 &  \ldots      &  \ldots       & -2.16$\pm$0.18&  5570$\pm^{260}_{250}  $ & $<$2.5\\
Garland-3 &   8.68 &8.68$\pm$0.20 &-1.21$\pm$0.14 &  \ldots      &  \ldots       & -2.28$\pm$0.17&  7430$\pm^{820}_{630}  $ & $<$2.5\\
Garland-4 &   8.75 &8.75$\pm$0.20 &-1.13$\pm$0.16 &  \ldots      &  \ldots       & -2.23$\pm$0.19&  7060$\pm^{840}_{660}  $ & $<$2.5 \\
Garland-6 &   8.89 &8.89$\pm$0.20 &-1.26$\pm$0.18 &  \ldots      &  \ldots       & -2.49$\pm$0.21&  5380$\pm^{570}_{430}  $ & $<$2.5\\
Garland-8 &   8.74 &8.74$\pm$0.20 &-1.18$\pm$0.15 &  \ldots      & \ldots & -2.34$\pm$0.19&  6720$\pm^{680}_{570}  $ & $<$2.5 \\
Garland-9 &   8.92 &8.92$\pm$0.20 &-1.35$\pm$0.15 &  \ldots      &  \ldots       & -2.24$\pm$0.19&  5140$\pm^{380}_{340}  $ & 2.9 \\

UGC 5666-1  &   7.87 &7.89$\pm$0.05 &-1.48$\pm$0.09 &-1.56$\pm$0.11& -0.83$\pm$0.12& -2.20$\pm$0.10& 13956$\pm^{990}_{890}  $ & 33\\
UGC 5666-2  &   7.81 &7.95$\pm$0.05 &-1.44$\pm$0.10 &  \ldots      & -0.81$\pm$0.15& -2.28$\pm$0.11& 12323$\pm^{940}_{910}  $ & 4.7\\
UGC 5666-3  &   7.90 &7.88$\pm$0.05 &-1.36$\pm$0.09 &-1.47$\pm$0.12& -0.75$\pm$0.13& -2.19$\pm$0.11& 14182$\pm^{1100}_{900} $ & 23\\
UGC 5666-5  &   7.99 &7.97$\pm$0.05 &-1.53$\pm$0.09 &-1.45$\pm$0.12& -0.80$\pm$0.13& -2.22$\pm$0.11& 13263$\pm^{940}_{840}  $ & 13\\
UGC 5666-6  &   7.93 &7.99$\pm$0.05 &-1.53$\pm$0.09 &-1.65$\pm$0.13& -0.80$\pm$0.13& -2.23$\pm$0.11& 12704$\pm^{850}_{810}  $ & 9.3\\
UGC 5666-8  &   7.87 &7.97$\pm$0.05 &-1.41$\pm$0.09 &-1.48$\pm$0.13& -0.81$\pm$0.14& -2.20$\pm$0.11& 12872$\pm^{1020}_{840} $ & 9.5\\
UGC 5666-10 &   7.84 &7.93$\pm$0.07 &-1.61$\pm$0.11 &  \ldots      & -0.65$\pm$0.18& -2.26$\pm$0.13& 13085$\pm^{1380}_{1230}$ & 4.6\\
UGC 5666-13 &   7.81 &7.83$\pm$0.05 &-1.43$\pm$0.09 &-1.51$\pm$0.12& -0.83$\pm$0.14& -2.23$\pm$0.11& 14158$\pm^{1220}_{1060}$ & 6.7\\
UGC 5666-9  &   7.81 &7.81$\pm$0.20 &-1.26$\pm$0.16 &-1.44$\pm$0.24& -0.68$\pm$0.31& -2.36$\pm$0.20& 15300$\pm^{3500}_{2600}$ & $<$2.5 \\
UGC 5666-11 &   7.86 &7.86$\pm$0.20 &-1.49$\pm$0.17 &-1.60$\pm$0.23& -0.78$\pm$0.30& -2.36$\pm$0.19& 13580$\pm^{2820}_{2030}$ & $<$2.5\\
UGC 5666-14 &   7.95 &7.95$\pm$0.20 &-1.53$\pm$0.17 &  \ldots      & -0.60$\pm$0.30& -2.41$\pm$0.19& 14430$\pm^{3020}_{2280}$ & $<$2.5\\

UGC 5692-3   &   7.98 &7.98$\pm$0.20 &-1.06$\pm$0.17 &  \ldots      &  \ldots       & -2.22$\pm$0.21& 12450$\pm^{3750}_{1750}$ & $<$2.5\\
UGC 5692-4   &   7.89 &7.89$\pm$0.20 &-0.99$\pm$0.18 &  \ldots      &  \ldots       & -2.02$\pm$0.21& 15800$\pm^{4200}_{3200}$ & $<$2.5\\
UGC 5692-6   &   7.99 &7.99$\pm$0.20 &-1.01$\pm$0.17 &  \ldots      &  \ldots       & -2.17$\pm$0.20& 12650$\pm^{2650}_{2050}$ & $<$2.5\\

UGC 5918-3   &   7.71 &7.84$\pm$0.04 &-1.23$\pm$0.07 &-1.41$\pm$0.10& -0.85$\pm$0.10& -2.05$\pm$0.09& 12905$\pm^{600}_{610}  $ & 6.1\\
UGC 5918-4   &   7.73 &7.73$\pm$0.20 &-1.53$\pm$0.16 &  \ldots      &  \ldots       & -2.03$\pm$0.18& 13820$\pm^{2480}_{2020}$ & $<$2.5 \\ 

UGC 8201-4  &   7.79 &7.79$\pm$0.20 & \nodata &  \ldots      & -0.78$\pm$0.51& \ldots & 15770$\pm^{4230}_{3170}$ & $<$2.5\\
UGC 8201-9  &   7.81 &7.81$\pm$0.20 & \ldots        &  \ldots      &  \ldots       &   \ldots      & 21400$\pm^{8600}_{5400}$ & $<$2.5
\enddata
\label{t:abund}
\tablenotetext{a}{Potential supernova remnant}
\end{deluxetable}

\begin{deluxetable}{ccccccccc}  
\tabletypesize{\scriptsize}
\tablewidth{0pt}
\tablecolumns{9}
\tablecaption{Gas Rich Galaxies of the M81 Group}
\tablehead{   
  \colhead{Galaxy}		&
  \colhead{Morphological}	&
  \colhead{Distance}		&
  \colhead{Distance}		&
  \colhead{M$_B$}		&
  \colhead{log(M$_{H I}$)}&
  \colhead{Log(L(H$\alpha$))}		&
  \colhead{12+log(O/H)}	&
  \colhead{References\tablenotemark{b}}	\\
  
  \colhead{}			&
  \colhead{Type}		&
  \colhead{(Mpc)}		&
  \colhead{Method\tablenotemark{a}}			&
  \colhead{}			&
  \colhead{(M$_\odot$)}	&
  \colhead{(ergs s$^{-1}$)} &
  \colhead{}			&
  \colhead{}		
}
\startdata
M 81		  	  & Sb	 & 3.63 &  Cep 	& -20.47$\pm$0.29 & 9.46 &40.77 & 9.00$\pm$0.13 & 1,2\\
M 82		  	  & Ir	 	 & 3.53 &  RGB & -19.37$\pm$0.30 & 8.90 &41.07 & 9.31$\pm$0.50 & 2\\
NGC 2403	  & Sc	 & 3.30 &  Cep 	& -18.85$\pm$0.33 & 9.56 &40.78 & 8.39$\pm$0.10 & 3\\
NGC 4236	  & Sd	 & 4.45 &  RGB & -18.24$\pm$0.30 & 9.46 &40.34 & 8.32$\pm$0.20 & 4\\
NGC 3077	  & Ir	 	 & 3.82 &  RGB & -17.74$\pm$0.30 & 8.80 &39.98 & 8.64$\pm$0.20 & 5\\
UGC 5666	  & Sm	 & 4.02 &  RGB & -17.34$\pm$0.30 & 9.23 &40.02 & 7.93$\pm$0.05 & 6,7,8\\
NGC 2976	  & Sm	 & 3.56 &  RGB & -17.05$\pm$0.30 & 8.28 &39.99 &\ldots&\\
UGC 4305	  & Im	 & 3.39 &  RGB & -16.70$\pm$0.30 & 8.99 &39.84 & 7.92$\pm$0.10 & 7,8\\
NGC 2366	  & Im	 & 3.19 &  RGB & -16.00$\pm$0.30 & 8.85 &40.10 & 7.91$\pm$0.05 & 7\\
UGC 8201	  & Im	 & 4.57 &  RGB & -15.09$\pm$0.30 & 8.14 &38.47 & 7.80$\pm$0.20 & 8\\
UGC 5692	  & Im	 & 4.00 &  RGB & -14.63$\pm$0.30 & $<$6.8 &38.47 & 7.95$\pm$0.20 & 8\\
UGC 5139	  & Ir	 	 & 3.84 &  RGB & -14.49$\pm$0.30 & 8.13 &38.82 & 8.00$\pm$0.10 & 6,8\\
UGC 5918	  & Ir	 	 & 7.40 &  BS    & -14.40$\pm$0.70 & 8.03 &38.64 & 7.84$\pm$0.04 & 8\\
UGC 7242	  & Ir	 	 & 5.42 &  RGB & -14.15$\pm$0.30 & 7.70 &$<$38.32 &\ldots&\\
UGC 6456	  & BCD  	 & 4.34 &  RGB & -14.03$\pm$0.30 & 7.80 &39.11 &7.73$\pm$0.05 & 9\\
UGC 5336	  & Ir	 	 & 3.70 &  MEM & -13.66$\pm$0.30 & 8.71 &38.19 & 8.65$\pm$0.25 & 8\\
UGC 4459	  & Ir	 	 & 3.56 &  RGB  & -13.37$\pm$0.30 & 7.62 &38.94 & 7.82$\pm$0.09 & 8\\
KDG 61		  & Sph? 	 & 3.60 &  RGB & -12.83$\pm$0.30 & $<$6.2 &37.83 &8.35$\pm$0.05 & 8\\
UGC 4483	  & BGC	 & 3.21 &  RGB & -12.73$\pm$0.30 & 7.52 &38.62 & 7.56$\pm$0.03 & 11,12\\
Garland  	  	  & Ir   	 & 3.55 &  RGB & -12.40$\pm$0.30 & 7.54 &39.05$^*$ &8.41$\pm$0.05 & 8\\
HS98 117 	  & Ir	 	 & 3.96 &  RGB & -11.98$\pm$0.30 & $<$5.0 &36.40$^*$ & 	\ldots &\\
A0952 		  & Ir	 	 & 3.87 &  RGB & -11.51$\pm$0.30 & 7.00 &38.00$^*$ & 	\ldots &\\
KDG 52		  & Ir	 	 & 3.55 &  RGB & -11.49$\pm$0.30 & 7.12 &$<$35.90 & 	\ldots &\\
KDG 73 		  & Ir	 	 & 3.70 &  RGB & -10.83$\pm$0.30 & 6.51 &$<$35.63 & 	\ldots &\\
BK 3N		  & Ir	 	 & 4.02 &  RGB & -9.59$\pm$0.30  & $<$6.5 &$<$35.64 & 	\ldots &
\enddata
\tablenotetext{a}{Cep = Cepheid Distance; RGB = Red Giant Branch fitting; BS = Brightest Star; MEM = Distance based on group membership}
\tablenotetext{b}{References for oxygen abundances are: (1) Zaritsky et al.\ 1994; (2) Alloin et al.\ 1979; (3) van Zee et al.\ 1998; 
(4) Matteucci \& Tosi 1985; 
(5) Storchi-Bergmann et al.\ 1994; (6) Miller \& Hodge 1996; (7) Masegosa et al.\ 1991; (8) this work; 
(9) Izotov et al.\ 1997; 
(10) Skillman et al.\ 1989; (11) van Zee \& Haynes 2006; (12) Skillman et al.\ 1994.}
\tablecomments{A listing of gas rich members of the M81 Group sorted in decreasing absolute blue magnitude.  Morphological type, distance, magnitude, and gas mass are from Karachentsev et al.\ (2002) with updates from Karachentsev et al.\ (2004).  H$\alpha$ luminosities are from Kennicutt et al.\ (2008) except those marked by an *, which are from Karachentsev \& Kaisin (2007).}
\label{t:m81}
\end{deluxetable}

\begin{figure}[htp] 
   \centering
   \plotone{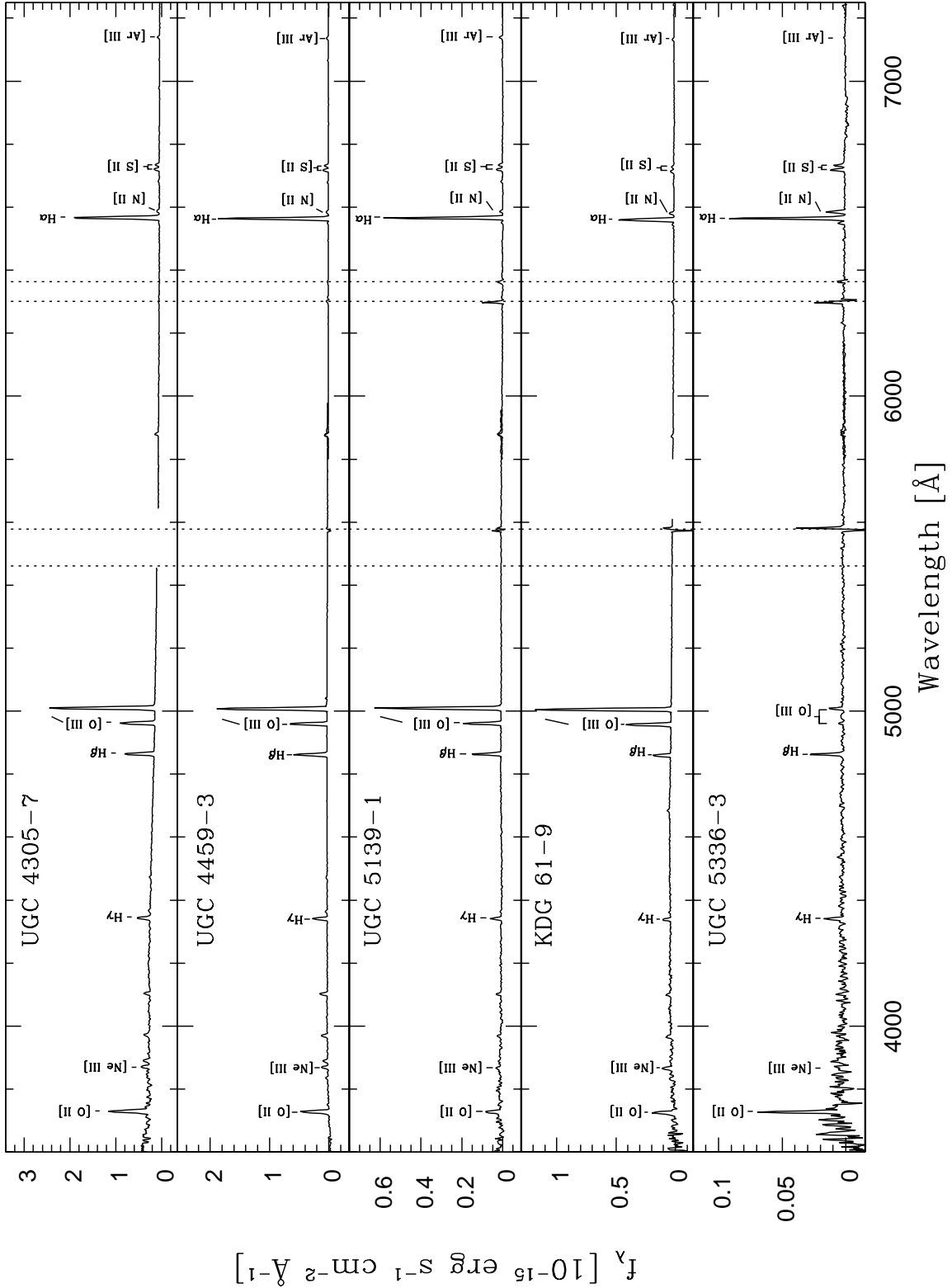}
   \caption{A representative spectrum for each target galaxy is shown.  The vertical dashed lines indicate the location of strong skylines.  }
   \label{fig:spec}
\end{figure}
\addtocounter{figure}{-1}
\begin{figure}[htp] 
   \centering
   \plotone{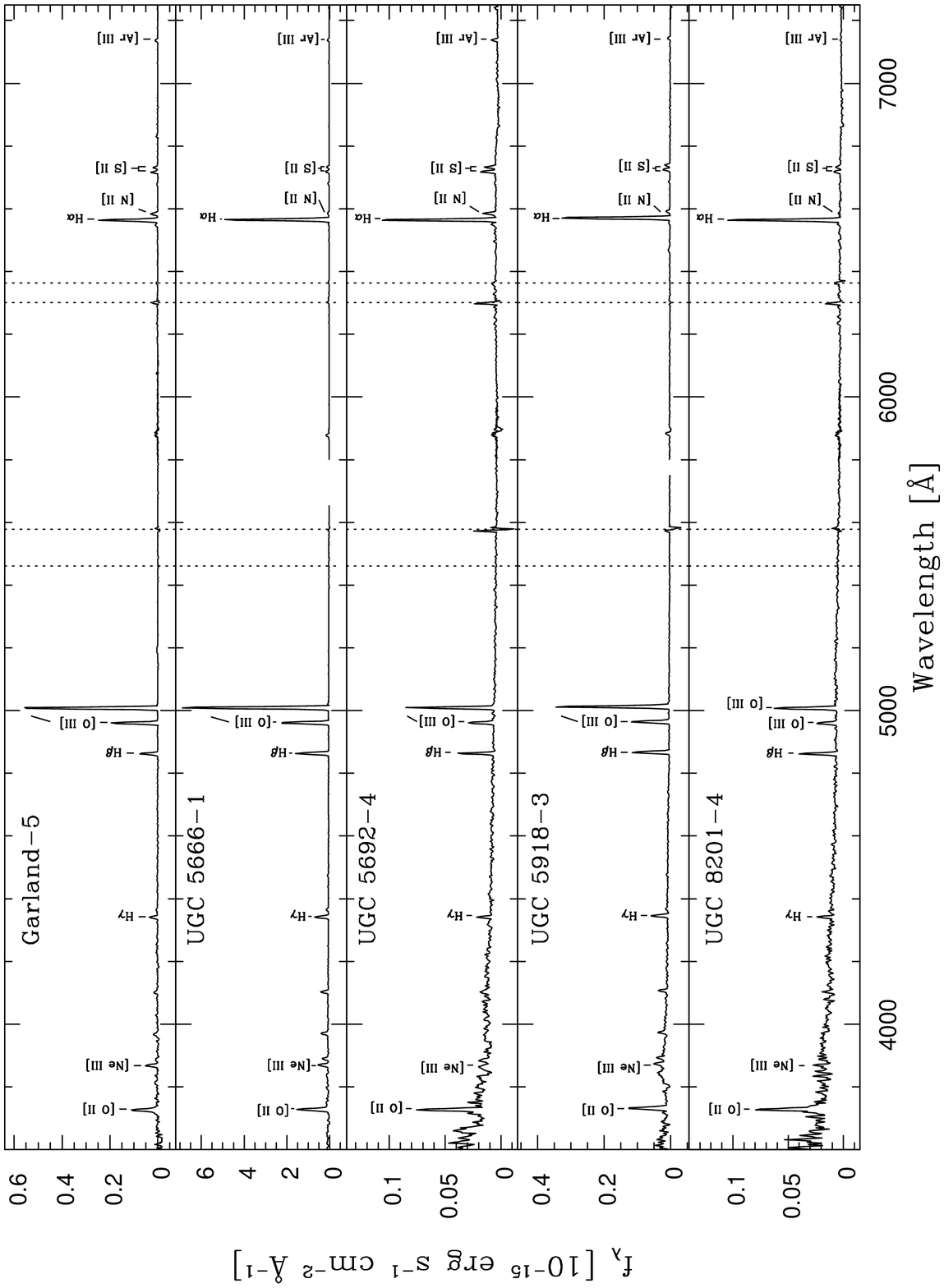}
   \caption{Continued}
\end{figure}

\begin{figure}[p] 
   \centering
   \plotone{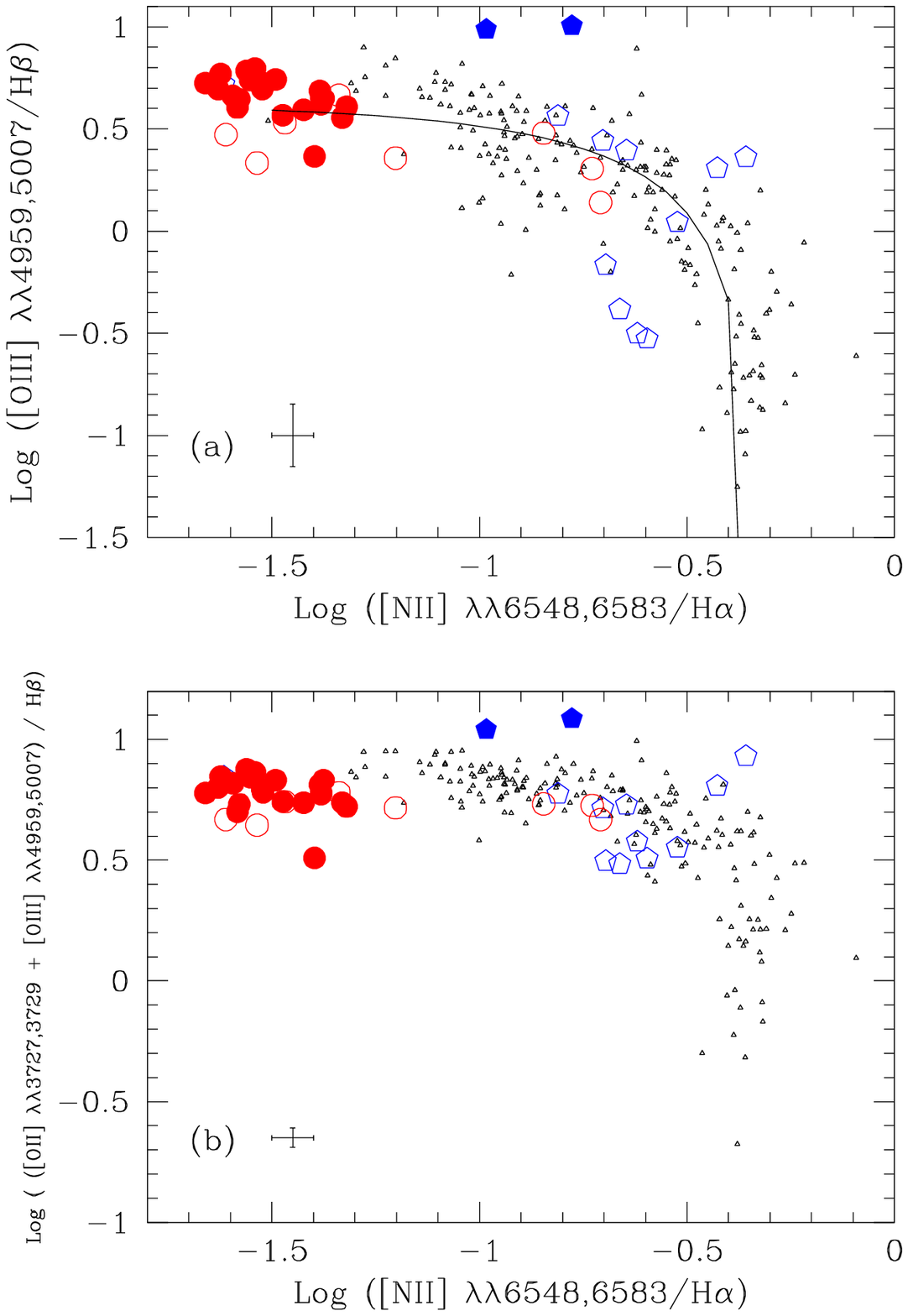}
   \caption{Emission-line diagnostic diagrams with [N~\ii ]/H$\alpha$.  Filled circles represent spectra in which the temperature sensitive [O~\iii] $\lambda$4363 line was detected; open circles and pentagons denote spectra where abundances were determined via the semi-empirical method.  Data from H~\ii\ regions in spiral galaxies (van Zee et al.\ 1998) are plotted as triangles to illustrate the scatter typical of diagnostics.  H~\ii\ regions from candidate tidal dwarfs, UGC 5336, KDG 61, and Garland are shown in blue petagons, while measurements from the remaining targets are shown in red circles (online only).  (a) The H~\ii\ region sequence formed by [N~\ii ] and [O~\iii] normalized by Balmer lines.  The theoretical prediction of Baldwin et al.\ (1981) is superposed.  (b) The H~\ii\ region sequence formed by R$_{23}$ versus [N~\ii ]/H$\alpha$.}
   \label{fig:diag}
\end{figure}

\begin{figure}[p] 
   \centering
   \plotone{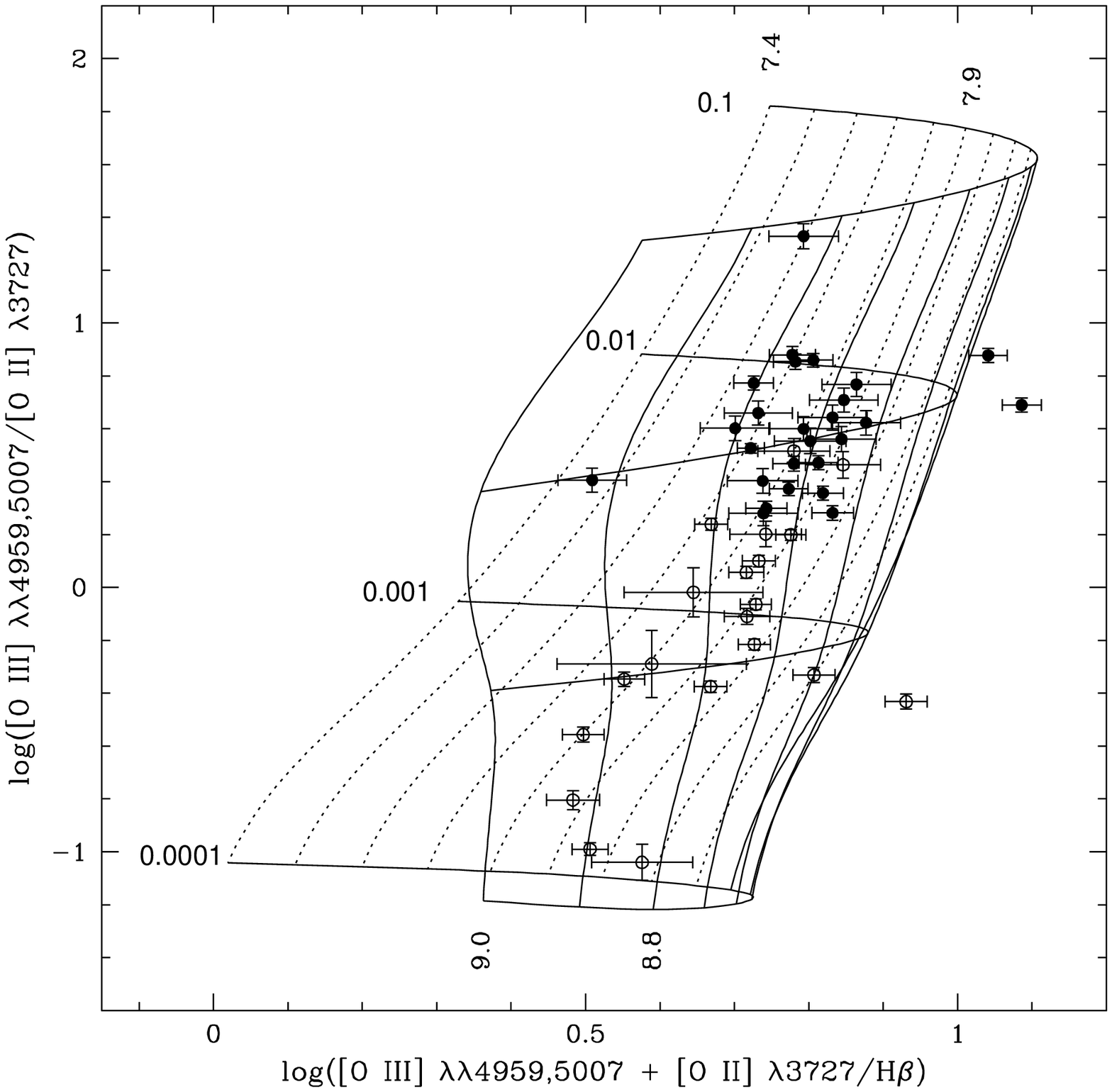}
   \caption{Model grid of the R$_{23}$ relation from McGaugh (1991).  The solid points indicate H~\ii\ regions where the temperature sensitive [O~\iii] $\lambda$4363 line was detected while open points denote H~\ii\ regions where this line was unmeasurable in the Gemini spectra.  The degeneracy between the high- and low-abundance regimes may be broken using the strength of the [N~\ii ] lines.}
   \label{fig:mcg}
\end{figure}

\begin{figure}[p] 
   \centering
   \plotone{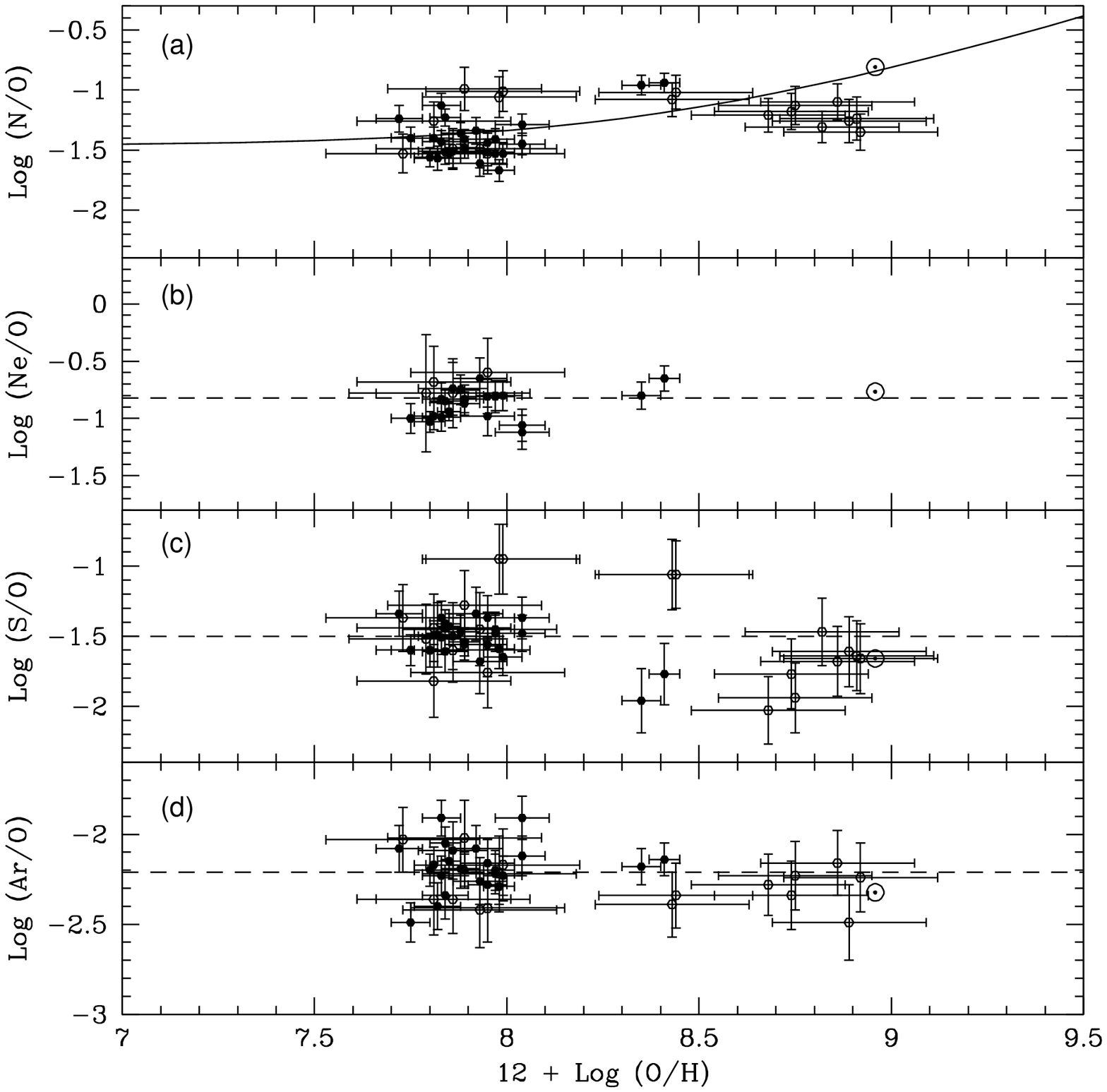}
   \caption{Relative enrichment of nitrogen and the $\alpha$ elements.  The solar value (Anders \& Grevesse 1989) is represented by $\odot$.  Filled circles represent spectra in which the temperature sensitive [O~\iii] $\lambda$4363 line was detected; open circles denote spectra where abundances were determined via the semi-empirical method.  The dashed line in panels (b), (c), and (d) denote the mean value. (a) Log (N/O) as a function of oxygen abundance.  The solid line designates the theoretical curve of Vila-Costas \& Edmunds (1993). (b) Log (Ne/O) as a function of oxygen abundance.  (c) Log (S/O) as a function of oxygen abundance.  (d) Log (Ar/O) as a function of oxygen abundance.}
   \label{fig:alpha}
\end{figure}

\begin{figure}[p] 
   \centering
   \plotone{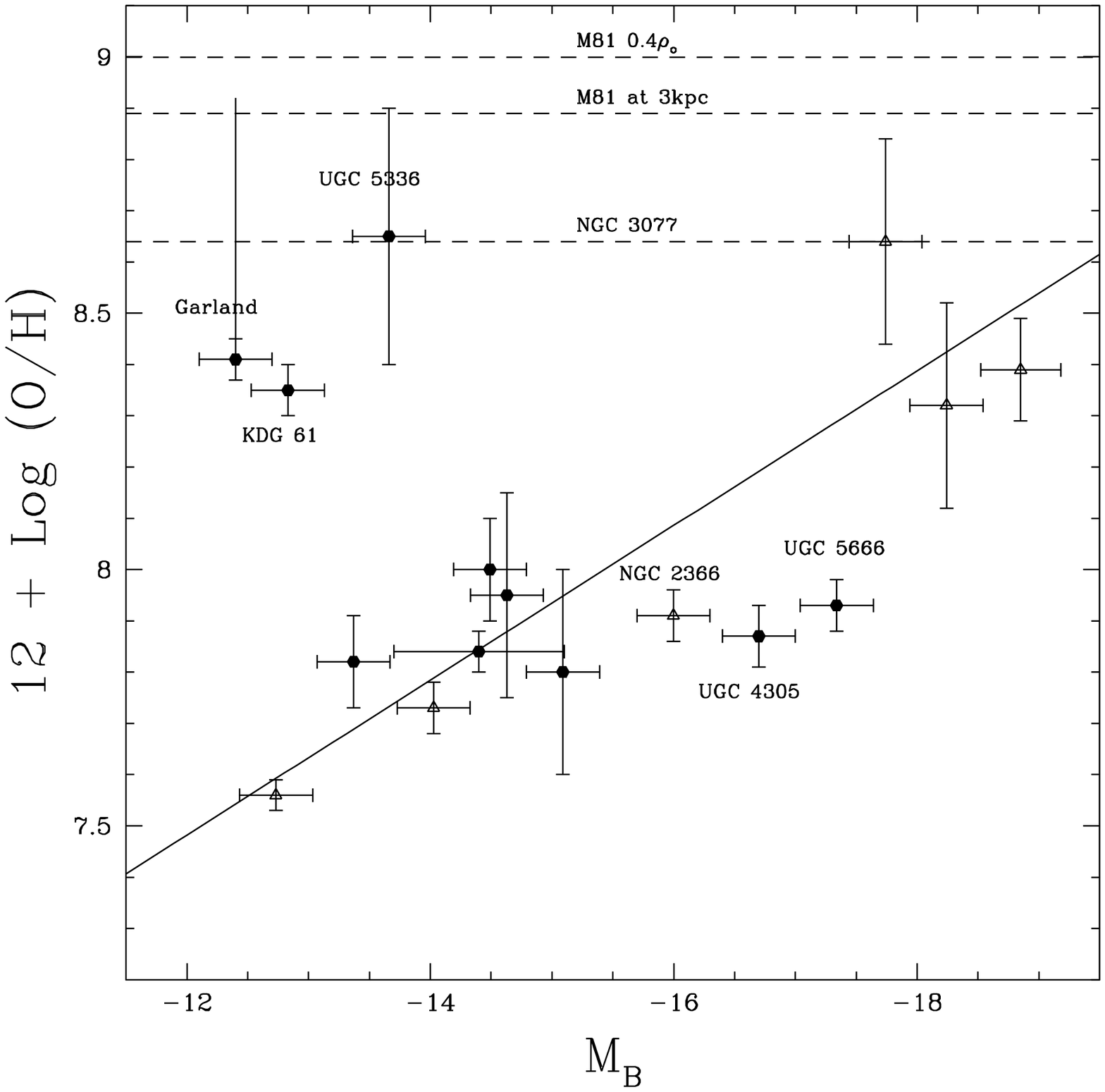}
   \caption{The metallicity-luminosity relation is shown for gas rich dwarfs belonging to the M81 Group.  Galaxies from this study are designated by filled circles, whereas triangles represent galaxies whose abundances were obtained from the literature (see Table~\ref{t:m81}). The long bar marking Garland indicates the spread of oxygen abundances measured. The solid line represents a least squares fit to oxygen abundances of field galaxies in the local volume (van Zee et al.\ 2006).  Abundances for M81 and NGC 3077 are marked by dashed lines (Zaritsky et al. 1994; Storchi-Bergman et al 1994, respectively).}
   \label{fig:metlum}
\end{figure}

\begin{figure}[p] 
   \centering
   \plotone{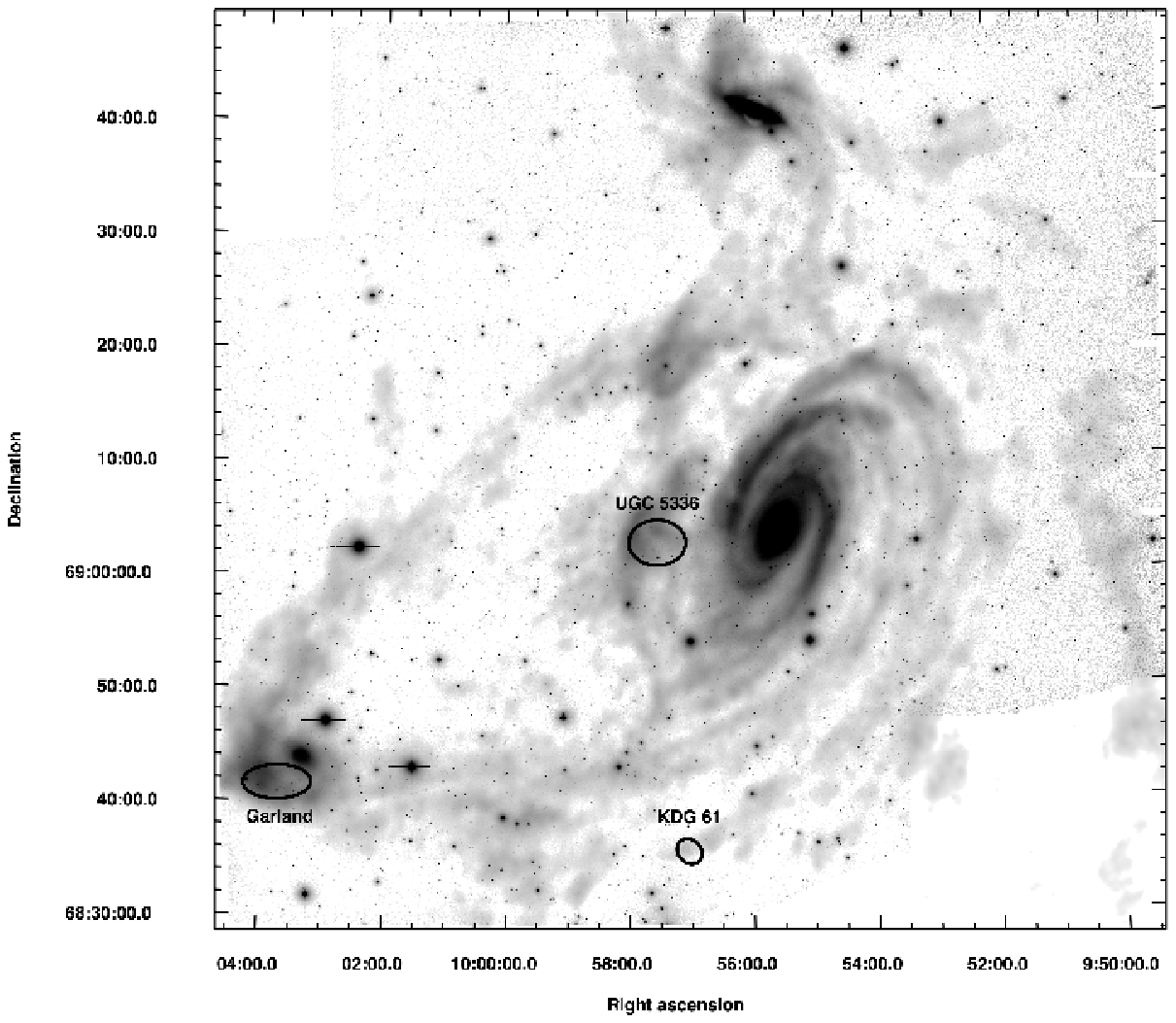}
   \caption{High-resolution H {\sc i} image of M81, M82 and NGC3077 and the tidal debris from their recent encounter (Yun et al.\ 1994) overlaid in blue on an I-band image of the same field. The locations and optical extent of UGC 5336, KDG 61, and Garland are indicated.  Note all three of these galaxies are coincident with tidal debris.}
   \label{fig:map}
\end{figure}

\begin{figure}[p] 
   \centering
   \plotone{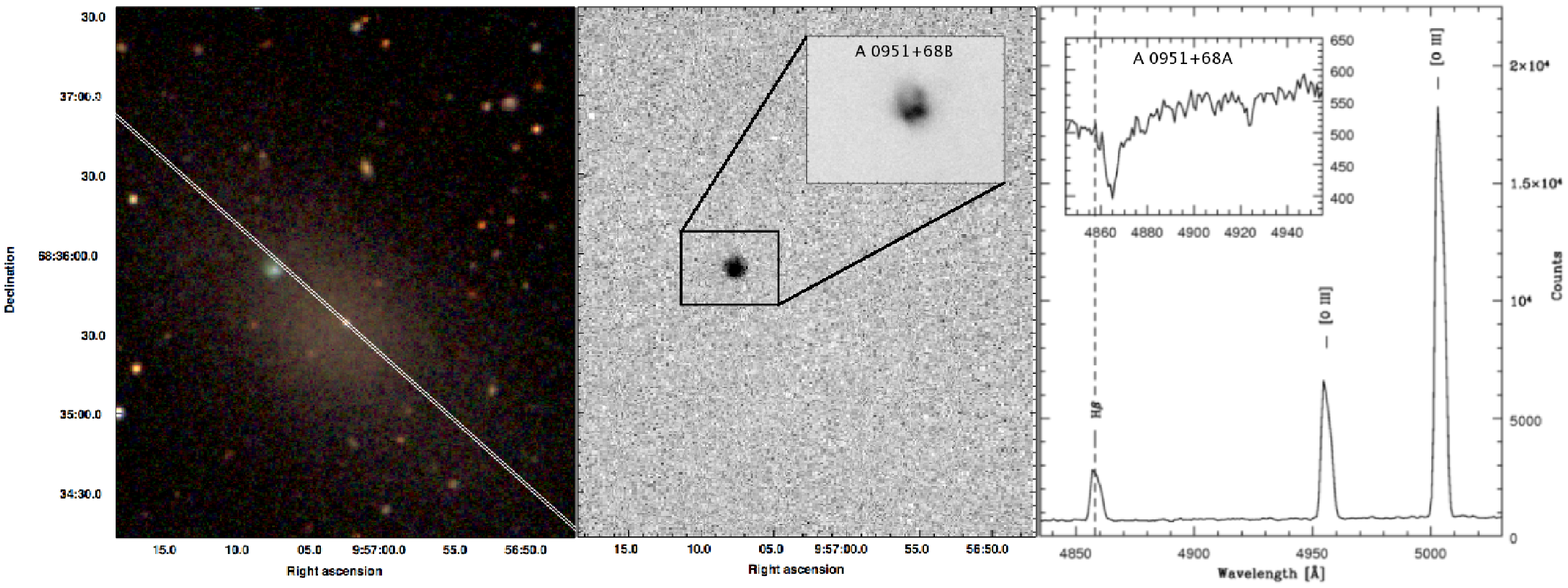}
   \caption{KDG 61: (Left) Three color optical image of KDG 61.  The slit location for the higher spectral resolution observations of Gemini Program GN-2006B-Q-29 is superposed.  The low spectral resolution observations of KDG 61-9 were centered on the bright blue knot north-east of the galaxy's center, A 0951+68B.  (Center) Continuum subtracted H$\alpha$ image of KDG 61. Note the bubble like structure visible in the enlarged inset. (Right) The H$\beta$ spectral region from the higher resolution Gemini long slit observations.  The vertical dashed line indicates the wavelength of H$\beta$ for an object moving with the M81 tidal stream at the location of KDG 61 (v~=~$-135$).  Emission lines from A 0951+68B have a redshift similar to that of the neutral gas.  In contrast, the central point-like object, A 0951+68A (shown in the inset), is not moving with the M81 tidal stream. }
   \label{fig:kdg}
\end{figure}
\end{document}